\documentclass[preprint,sort&compress,12pt]{elsarticle}

\usepackage{graphicx}
\usepackage{amsmath}

\usepackage{natbib}
\usepackage{amssymb}

\usepackage{amsthm}

\usepackage{lineno}
\usepackage{subfig}
\usepackage{enumerate}
\usepackage{fullpage}
\usepackage{algorithm}
\usepackage{algpseudocode}
\usepackage{xcolor}
\usepackage[colorinlistoftodos]{todonotes}

\biboptions{sort,compress} 

\usepackage{xcolor}

\newcommand*\patchAmsMathEnvironmentForLineno[1]{%
  \expandafter\let\csname old#1\expandafter\endcsname\csname #1\endcsname
  \expandafter\let\csname oldend#1\expandafter\endcsname\csname end#1\endcsname
  \renewenvironment{#1}%
     {\linenomath\csname old#1\endcsname}%
     {\csname oldend#1\endcsname\endlinenomath}}%
\newcommand*\patchBothAmsMathEnvironmentsForLineno[1]{%
  \patchAmsMathEnvironmentForLineno{#1}%
  \patchAmsMathEnvironmentForLineno{#1*}}%
\AtBeginDocument{%
\patchBothAmsMathEnvironmentsForLineno{equation}%
\patchBothAmsMathEnvironmentsForLineno{align}%
\patchBothAmsMathEnvironmentsForLineno{flalign}%
\patchBothAmsMathEnvironmentsForLineno{alignat}%
\patchBothAmsMathEnvironmentsForLineno{gather}%
\patchBothAmsMathEnvironmentsForLineno{multline}%
}

\usepackage{color,soul}
\usepackage{enumitem}
\usepackage{mathtools}
\usepackage{booktabs}

\definecolor{lightblue}{rgb}{.90,.95,1}
\definecolor{darkgreen}{rgb}{0,.5,0.5}

\definecolor{lightgreen}{rgb}{.90,1,0.90}

\newcommand{\bstau}{\boldsymbol{\tau}}
\newcommand{\bstaurans}{\tilde{\boldsymbol{\tau}}^{rans}}

\graphicspath{ {./figs/} {./figs/jinlong/} {./figs/jianxun/}}

\journal{Springer }

\begin{document}

\begin{frontmatter}



\title{Quantifying Model Form Uncertainty in RANS Simulation of Wing--Body Junction Flow}



\author{Jin-Long Wu\corref{corjl}}
\author{Jian-Xun Wang\corref{corjxw}}
\author{Heng Xiao\corref{corxh}}
\cortext[corxh]{Corresponding author. Tel: +1 540 231 0926}
\ead{hengxiao@vt.edu}

\address{Department of Aerospace and Ocean Engineering, Virginia Tech, Blacksburg, VA 24060, United States}

\begin{abstract}
Wing--body junction flows occur when a boundary layer encounters an airfoil mounted on the surface. It is characterized by a horseshoe vortex that arises at the upstream of the leading edge of airfoil and a possible corner separation that develops within the junction. The corner flow near the trailing edge is challenging for the linear eddy viscosity Reynolds Averaged Navier-Stokes (RANS) models, due to the interaction of two perpendicular boundary layers which leads to highly anisotropic Reynolds stress at the near wall region. Recently, Xiao et al. (Xiao, Wu, Wang, Sun, and Roy. Quantifying and reducing model-form uncertainties in Reynolds averaged Navier-Stokes equations: An open-box, physics-based, Bayesian approach. Submitted. arxiv preprint number: 1508.06315) proposed a physics-informed Bayesian framework to quantify and reduce the model-form uncertainties in RANS simulations by utilizing sparse observation data. In this work, we extend this framework to incorporate the use of wall function in RANS simulations, and apply the extended framework to the RANS simulation of wing--body junction flow. Standard RANS simulations are performed on a 3:2 elliptic nose and NACA0020 tail cylinder joined at their maximum thickness location (known as 'Rood' wing). The uncertainties are directly injected into the RANS predicted Reynolds stresses under the constrain of turbulence realizability, and the observation data are then incorporated by using the iterative ensemble Kalman method. Current results show that both the posterior mean velocity and the Reynolds stress anisotropy show better agreement with the experimental data at the corner region near the trailing edge, which demonstrates the capability of this framework in improving the RANS simulation of the wing--body junction flow. On the other hand, the prior velocity profiles at the leading edge indicates the restriction of uncertainty space and the performance of the framework at this region is less effective. By perturbing the orientation of Reynolds stress, the uncertainty range of prior velocity profiles at the leading edge covers the experimental data. It indicates that the uncertainty of RANS predicted velocity field is more related to the uncertainty in the orientation of Reynolds stress at the region with rapid change of mean strain rate. The present work not only demonstrates the capability of Bayesian framework in improving the RANS simulation of wing--body junction flow, but also reveals the major source of model-form uncertainty for this flow, which can be useful in assisting RANS modeling.

\end{abstract}

\end{frontmatter}


\section{Introduction}
\label{sec:intro}
Wing--body junction flows are common in many engineering applications, such as the airfoil/fuselage junction of an aircraft, the sail/hull junction of a submarine and the blade/hub assembly of a wind turbine. Due to the flow stagnation at the leading edge of the wing, strong streamwise adverse pressure gradient occurs at upstream of the stagnation point. Such adverse pressure gradient leads to the separation of the incoming boundary layer, which forms spanwise vortex upstream of the leading edge. This vortex will move around the wing and stretch along the streamwise direction, which is known as the horseshoe vortex system. The formation of this vortex system has been well studied for decades~\cite{devenport1990,simpson2001}. On the other hand, the flow near the trailing edge of the junction is much more complex, due to the combination effect of vortex stretching, adverse pressure gradient and the interaction between the boundary layer on the wing and that on the body. This interaction of two perpendicular boundary layers leads to the stress-induced secondary flow, which cannot be predicted by the linear eddy viscosity RANS models, including $k$--$\varepsilon$ model, $k$--$\omega$ model and S--A model. This is because these linear viscosity RANS models have difficulty in predicting the anisotropy state of Reynolds stresses at the near wall region. Although the nonlinear eddy viscosity models are able to predict the stress-induced secondary flow, these models are still restricted by the assumption that the Reynolds stress is determined by the local mean flow quantities. Compared to RANS simulations, high fidelity simulations such as Large Eddy Simulation (LES) and Direct Numerical Simulation (DNS) are more accurate for the prediction of such stress-induced secondary flow. However, the computational cost of these high fidelity simulations is currently prohibitive for most real engineering applications of junction flow, since the wall bounded flow at high Reynolds number demands fine resolution of the near wall region.

Many previous studies of junction flow have focused on the horseshoe vortex system. Rodi et al.~\cite{rodi1998} compared the algebraic Reynolds stress model and the non-linear $k$--$\varepsilon$ models with the linear eddy viscosity models, and showed that the more complex RANS models were no better than linear eddy viscosity models for the practical mean quantities such as mean velocity. Coombs at al.~\cite{coombs2012} tested eight turbulence models including RNG $k$--$\varepsilon$ model and LRR model  and found that all models significantly under-predicted the turbulent kinetic energy (TKE). Aspley and Leschziner~\cite{apsley2001} compared twelve RANS models including non-linear eddy viscosity models and Reynolds stress transport models (RSTM), and suggested that the second-moment closure models offered better prediction of Reynolds stress over other models, although no model achieved close agreement with the experimental data. Similarly, Chen~\cite{chen1995} showed that RSTM had an overall better prediction than the isotropic eddy viscosity models. However, the convergence could be difficult to achieve for RSTM~\cite{jones2005}. All these previous studies indicate that further increasing the complexity of RANS model barely improve the prediction of mean quantities that is of the most interest in engineering applications.

The hybrid RANS/LES simulations, such as Detached Eddy Simulations (DES), have been used in several previous studies and provide better prediction for the horseshoe vortex system. However, the predicted location of the horseshoe vortex system is still not satisfactory. Alin and Fureby~\cite{alin2008} showed that the DES typically predicts the horseshoe vortex located too further away from the body than the experimental measurements. Paik et al.~\cite{paik2007} also reported the discrepancy in predicting the location of horseshoe vortex system. In addition, Paik et al.~\cite{paik2007} pointed out that the DES simulation result depends closely on the flow-specific adjustment of the DES length scale. 

Compared to the horseshoe vortex system at the leading edge, the flow separation at the trailing edge of junction is much less investigated. Huser and Biringen~\cite{huser1993direct} showed that the normal stress imbalance was important for the generation of stress-induced secondary flow within the junction of two flat plates. For a more realistic configuration with NACA0012 airfoil,  Gand et al.~\cite{gand2012} found that the prediction of flow separation at the trailing edge varied based on different linear eddy viscosity RANS models, and none of the prediction had a close agreement with the experimental measurement. In addition, Gand et al.~\cite{gand2012,gand2015} also pointed out that the linear eddy viscosity models were not able to accurately predict the corner separation of junction flow. Bordji et al.~\cite{bordji2014} showed that the quadratic constitutive relation (QCR) closure can provide better prediction of corner separation, and they suggested that the effect of corner flow separation may also be associated to the local increase of turbulent kinetic energy. Rumsey et al.~\cite{rumsey2016} confirmed that the QCR closure improved the prediction of corner separation compared to the linear model. However, he also pointed out that the existing comparisons were not all consistent and more efforts were still required in understanding the corner separation of wing--body junction flow. 

Recently, Xiao et al.~\cite{xiao-mfu} proposed a physics-informed, data-driven Bayesian framework for calibrating the RANS simulations by quantifying and reducing the model-form uncertainties in RANS simulations with a small amount of velocity observation data. In their framework, uncertainties were introduced to the Reynolds stresses and a Bayesian inference procedure based on an iterative ensemble Kalman method~\cite{iglesias2013ensemble} was used to quantify and reduce the uncertainties by incorporating observation data. By applying their framework to the RANS simulation of the flow in a square duct, they had demonstrated that the prediction of normal stress imbalance can be improved and the secondary flow was therefore captured. 

However, to avoid the possible complexity caused by the wall functions, wall resolved RANS simulations were used in the framework by Xiao~\cite{xiao-mfu}, which demand large computational cost for wall bounded flow at high Reynolds number. In practical scenarios, many RANS simulations are performed with wall functions to make the computational cost affordable. Therefore, in the present work we explore the compatibility of this Bayesian framework and the wall function approach commonly used in RANS simulations. Specifically, we discuss how the wall function is incorporated into the RANS simulation, and demonstrate how the wall function is taken into consideration in our current framework. To illustrate this concept, we use the implementation of OpenFOAM as an example. Based on this extension, we are able to apply the Bayesian framework to quantify and reduce the model-form uncertainty in RANS simulations with complex geometry at higher Reynolds numbers. 


In addition to the compatibility with wall function, the arrangement of observation is also discussed in this work. In the original framework proposed by Xiao et al.~\cite{xiao-mfu}, the arrangement of observation is determined based on the physical understanding of the particular flow. Specifically, it is determined based on an empirical estimation of the mean flow correlation. Since such correlation is important for the performance of Bayesian inference with sparse observation data, it is more rigorous to arrange the observations with an objective criteria rather than the subjective judgement by the user, especially for the complex flow problem in which the mean flow correlation is less obvious. In this work, we apply the correlation analysis to the mean velocity field to estimate the mean flow correlation, and determine the arrangement of observation accordingly.

The objective of the present work is to use the Bayesian framework proposed by Xiao et al.~\cite{xiao-mfu} to improve the prediction of the RANS simulation of the wing--body junction flow. This is a much more complicated flow problem compared to the flow problem that studied in the work by Xiao et al.~\cite{xiao-mfu}. We first extend the original Bayesian framework by Xiao et al.~\cite{xiao-mfu} for complex wall-bounded flows at high Reynolds numbers. Based on the extended framework, we further calibrate the RANS simulation of wing--body junction flow, and demonstrate that the predicted mean flow field by the linear eddy viscosity RANS model can be improved. The remaining of the paper is organized as follows. Section 2 outlines the Bayesian framework proposed by Xiao et al.~\cite{xiao-mfu} and presents the extension in the current work. Section 3 presents the numerical simulation results of wing--body junction flow and compares the results with the experimental data~\cite{devenport1990}. Section 4 discusses the limitation of the current framework and the possible extension. Finally, Section 5 concludes the paper.

\section{Methodology of the Model-Form Uncertainty Quantification Framework}
\label{sec:method}

\subsection{Summary of the Model-Form Uncertainty Quantification Framework}

\begin{figure}[!htbp]
  \centering
   \subfloat[Barycentric coordinate]
   {\includegraphics[width=0.5\textwidth]{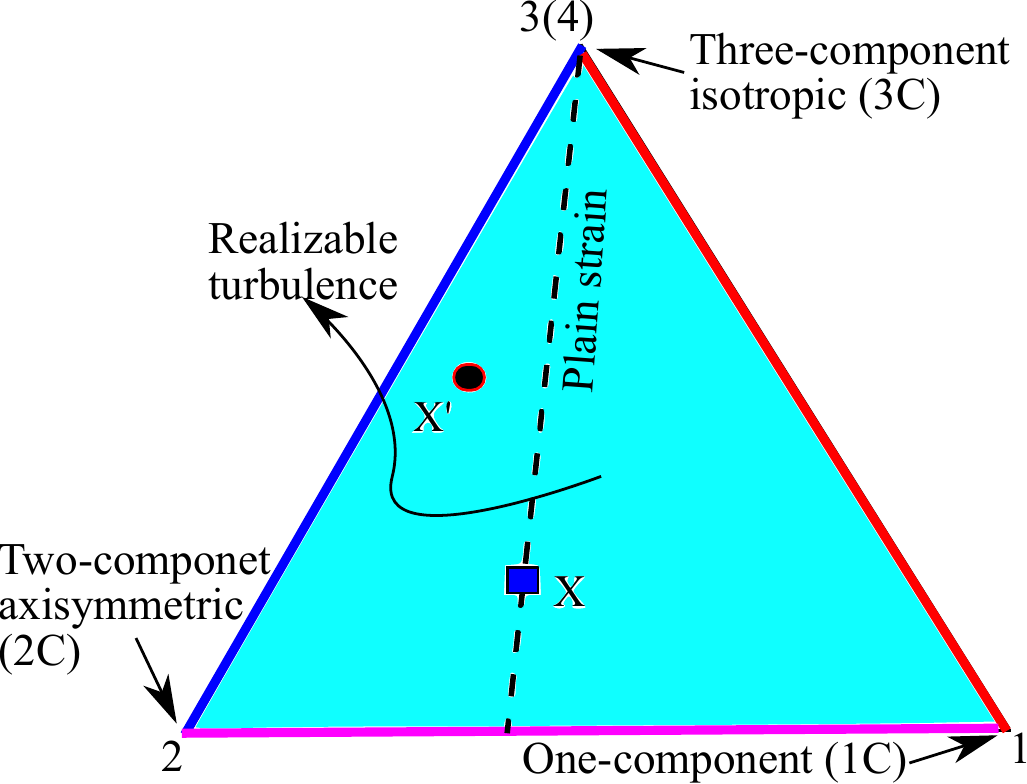}} \hspace{2em}
   \subfloat[natural coordinate]
   {\includegraphics[width=0.4\textwidth]{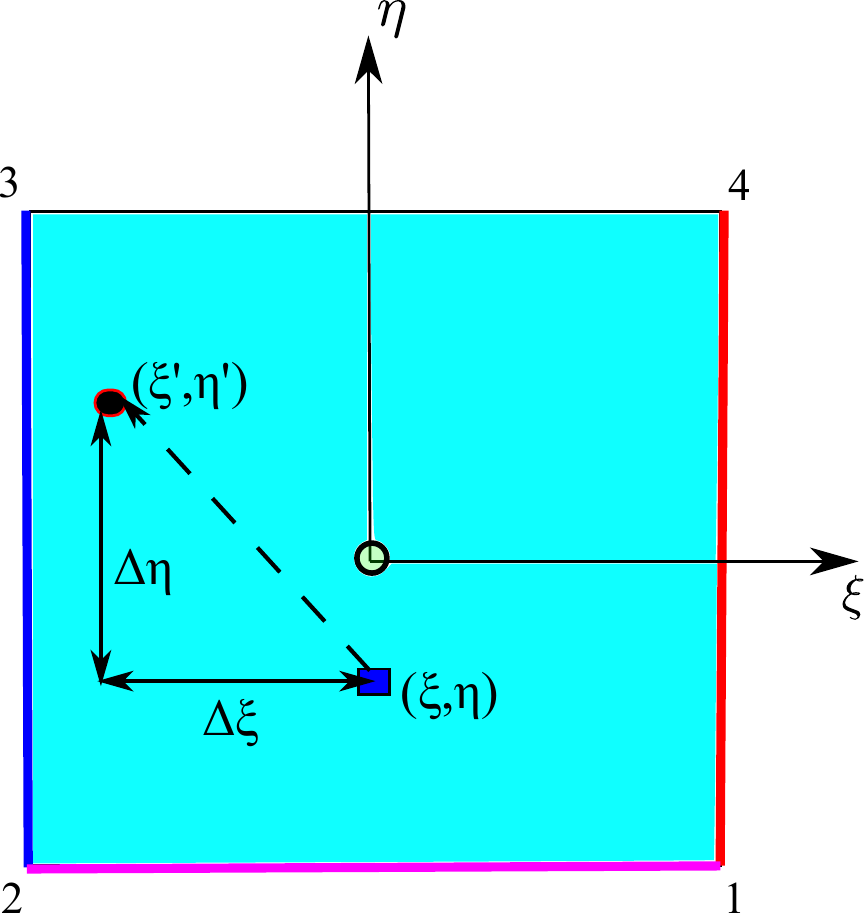}}\\
   \caption{Mapping between the Barycentric coordinate to the natural coordinate, transforming the
     Barycentric triangle enclosing all physically realizable
     states~\cite{banerjee2007presentation,emory2013modeling} to a square via standard finite
     element shape functions. Corresponding edges in the two
     coordinates are indicated with matching colors.}
  \label{fig:bary}
\end{figure}

We first summarize the Bayesian framework, which is proposed by Xiao et al.~\cite{xiao-mfu} for quantifying and reducing model-form uncertainties in RANS simulations. In RANS simulations, the modeled Reynolds stress term is considered as the main source of model-form uncertainty~\cite{oliver2011bayesian}. To quantify this model-form uncertainty, perturbations are directly injected to the RANS-modeled Reynolds stress. Specifically, the Reynolds stress $\bstau(x)$ term is modeled as a random field whose prior mean is the RANS-predicted Reynolds stress $\bstaurans(x)$, in which $x$ represents the spatial coordinates. It should be noted that Reynolds stress tensor is a positive semi-definite matrix, which leads to the realizability requirement of Reynolds stress~\cite{tennekes1972first}. This realizability requirement may be violated if arbitrary perturbation is injected into each element of the Reynolds stress tensor. To guarantee the realizability of Reynolds stress, Iaccarino and co-workers~\cite{gorle2013framework,gorle2014deviation,emory2013modeling,emory2011modeling,emory14estimate} first proposed an uncertainty quantification approach to perform the perturbation within the Barycentric triangle, which guarantees the realizability of Reynolds stress. In details, the Reynolds stress tensor is decomposed into physical meaningful components and transformed into the coordinates of Barycentric triangle~\cite{banerjee2007presentation} as follows:

\begin{equation}
  \label{eq:tau-decomp}
  \boldsymbol{\tau} = 2 k \left( \frac{1}{3} \mathbf{I} +  \mathbf{a} \right)
  = 2 k \left( \frac{1}{3} \mathbf{I} + \mathbf{V} \Lambda \mathbf{V}^T \right)
\end{equation}
where $k$ is the turbulent kinetic energy, which indicates the magnitude of $\bstau$; $\mathbf{I}$
is the second order identity tensor; $\mathbf{a}$ is the anisotropy tensor; 
$\mathbf{V} = [\mathbf{v}_1, \mathbf{v}_2, \mathbf{v}_3]$ and 
$\Lambda = \textrm{diag}[\lambda_1, \lambda_2, \lambda_3]$ with
$\lambda_1+\lambda_2+\lambda_3=0$ are the eigenvectors and eigenvalues of $\mathbf{a}$,
respectively. The eigenvalues $\lambda_1$,
$\lambda_2$, and $\lambda_3$ are mapped to a Barycentric coordinate $(C_1, C_2, C_3)$ with $C_1 +
C_2 + C_3 = 1$. Consequently, the Barycentric
triangle shown in Fig.~\ref{fig:bary}a encloses all physically realizable states of Reynolds stress. To facilitate the parameterization, the Barycentric
coordinate is further transformed to the natural coordinate $(\xi, \eta)$ as shown in Fig.~\ref{fig:bary}b.  Finally, uncertainties are introduced to the mapped
quantities $k$, $\xi$, and $\eta$ by adding discrepancy terms to the corresponding RANS predictions,
i.e.,
\begin{subequations}
    \label{eq:delta-def}
  \begin{alignat}{2}
    \log k(x) & = &\ \log \tilde{k}^{rans}(x)  & + \delta^k(x)  \label{eq:kdelta} \\
    \xi (x) & = &\ \tilde{\xi}^{rans}(x) & + \delta^\xi(x)  \\
    \eta(x) & = &\ \tilde{\eta}^{rans}(x) & + \delta^\eta(x)
  \end{alignat}
\end{subequations}
Currently, uncertainties are not introduced to the orientation ($\mathbf{v}_1, \mathbf{v}_2, \mathbf{v}_3$) of
the Reynolds stress in the Bayesian inference framework. This is due to the consideration of numerical stability. More details has been discussed in~\cite{wu2015bayesian}.

The prior discrepancies are chosen as nonstationary
zero-mean Gaussian random fields $\mathcal{GP}(0, K)$ (also known as Gaussian processes), and the
basis set is chosen as the eigenfunctions of the kernel $K$~\cite{le2010spectral}. Therefore, 
the discrepancies can be reconstructed as follows:
\begin{equation}
  \label{eq:delta-proj}
  \delta(x) = \sum_{i=1}^\infty \omega_{i} \; \phi_i (x) , 
\end{equation}
where $\phi_i (x)$ are the eigenfunctions of the kernel $K$, and the coefficients $\omega_{i}$  are independent standard Gaussian random variables. In practice, the
infinite series is truncated to $m$ terms with $m$ depending on the smoothness of the kernel $K$.

With the projections above, the discrepancies are parameterized by the coefficients $\omega^k_{i},
\, \omega^\xi_{i}, \, \omega^\eta_{i}$ with $i = 1, 2, \cdots, m$. These coefficients are then inferred by an iterative ensemble Kalman
method~\cite{iglesias2013ensemble,evensen2009data}. Specifically, this iterative ensemble Kalman
method incorporates sparse observation data of mean velocity from experiment to infer the
coefficients $\omega_{i}$ of the Reynolds stress discrepancies as shown in
Eq.~\ref{eq:delta-proj}. More details of the Bayesian inference procedure can be found in ref.~\cite{xiao-mfu}. 

\subsection{Compatibility of the Uncertainty Quantification Framework with Wall Function}

The original Bayesian framework proposed by Xiao et al.~\cite{xiao-mfu} is developed for wall-resolved RANS simulations and is restricted for many engineering applications, since resolving the near wall region is costly for wall-bounded flow at high Reynolds number. In this work, we discuss the compatibility of this Bayesian framework for RANS simulations with wall functions. For the wall resolved RANS simulation, the wall shear stress can be estimated as follow:
\begin{equation}
  \label{eq:tauw_resolved}
  \tau_w = \nu \frac{U_{p}}{y_p}
\end{equation}
where $U_{p}$ represents the mean velocity tangential to the wall at the first cell near the wall, and $y_p$ is the distance between the wall and the center of that cell. 
However, Eq.~\ref{eq:tauw_resolved} is only a valid approximation for wall shear stress if the viscous sublayer is resolved, which demands much higher computational cost than RANS simulation with wall functions. In the practice of RANS simulation with wall functions, the center of first cell is arranged in the logarithmic layer. Therefore, Eq.~\ref{eq:tauw_resolved} is no longer appropriate to estimate the wall shear stress. Instead, the wall shear stress is estimated at the first layer of cell near wall as follow:
\begin{equation}
  \label{eq:tauw}
  \tau_w = (\nu_t+\nu) \frac{U_{p}}{y_p}
\end{equation}
where $\nu_t$ is calculated as follow:
\begin{equation}
  \label{eq:nut}
  \nu_t = \nu \left(\frac{y^+ \kappa}{\log{(E y^+)}} -1\right)
\end{equation}
Essentially, the estimation of wall shear stress based on Eq.~\ref{eq:tauw} and Eq.~\ref{eq:nut} makes use of the law of the wall. By substituting Eq.~\ref{eq:nut} into Eq.~\ref{eq:tauw} and recalling that $u_{\tau}=\sqrt{\tau_w / \rho}$, $y^+=y \sqrt{\tau_w} / \nu$ and $u^+=U / u_{\tau}$, we can obtain the log law of the wall:
\begin{equation}
  \label{eq:law_wall_2}
  u^+=\frac{1}{\kappa}\log{y^+} + \frac{1}{\kappa} \log{E}
\end{equation}
where the constant $E$ is 9.8~\cite{popebook} and the constant term $C^+=\frac{1}{\kappa} \log{E} \approx$ 5. It should be noted that the true value of $y^+$ is unknown before the wall shear stress $\tau_w$ is solved. In practice, $y^*$ as calculated by Eq.~\ref{eq:ystar} is used as an approximation of $y^+$~\cite{launder1974numerical}.
\begin{equation}
  \label{eq:ystar}
  y^*=\frac{\left( C_{\mu}^{1/2} k_p \right)^{1/2}}{\nu}
\end{equation}
where $k_p$ represents the turbulent kinetic energy at the first cell near the wall.

The above derivation shows that the wall shear stress calculated by Eq.~\ref{eq:tauw} is compatible with the law of the wall, by specifying the boundary condition of eddy viscosity $\nu_t$ at wall based on Eq.~\ref{eq:nut}. To illustrate how we incorporate the wall functions in our current uncertainty quantification framework, we use the implementation in OpenFOAM as an illustration. Compared to the built-in solvers of OpenFOAM, the CFD solver tauFOAM used in the Bayesian framework directly takes the Reynolds stress field as an input. Theoretically, each component of Reynolds stress at wall should be zero, due to the non-slip condition. However, if we specify all the components of Reynolds stress as zero at wall, the wall shear stress turns to Eq.~\ref{eq:tauw_resolved}, which underestimates the wall shear stress if the near wall mesh is coarse and the viscous sublayer is not resolved. In such case, to estimate the wall shear stress based on the law of wall, we first calculate the value of $\nu_t$ at wall based on Eq.~\ref{eq:nut}. In our uncertainty quantification framework, the Reynolds stress boundary condition at wall is specified according to Eq.~\ref{eq:tau_R} if the viscous sublayer is not resolved:
\begin{equation}
  \label{eq:tau_R}
  \widetilde{\tau}_w = \nu_t \frac{U_{p}}{y_p}
\end{equation}
The wall shear stress is calculated as follow:
\begin{equation}
  \label{eq:tauw3}
  \tau_w = \widetilde{\tau}_w+\nu \frac{U_{p}}{y_p}
\end{equation}
which is essentially the same as Eq.~\ref{eq:tauw}. Therefore, the wall shear stress is still compatible with the law of the wall as shown in Eq.~\ref{eq:law_wall_2} due to the specification of boundary condition of Reynolds stress. 

\section{Numerical Simulations}
\label{sec:simulations}

\subsection{Case Setup}

\begin{figure}[htbp]
  \centering
  \includegraphics[width=0.75\textwidth]{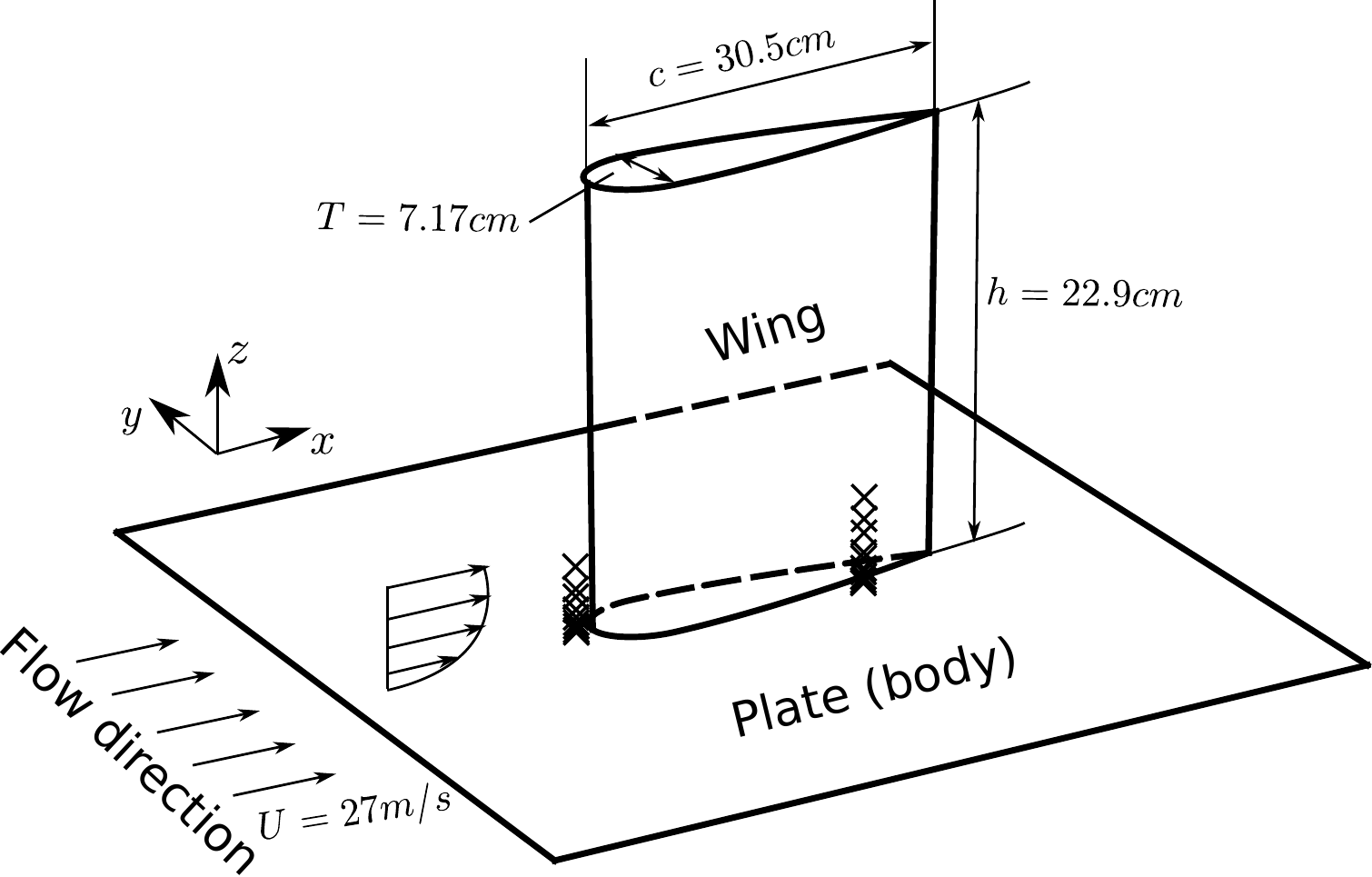}
  \caption{Domain shape of the wing--body junction flow. The $x$-, $y$- and $z$-coordinates are
    aligned with streamwise, spanwise of the plate and spanwise of the airfoil, respectively. The locations where velocities
    are observed are indicated as crosses ($\times$).}
  \label{fig:domain_junction}
\end{figure}

The configuration of the computational domain and the coordinate system are shown in Fig.~\ref{fig:domain_junction}. The airfoil shown in Fig.~\ref{fig:domain_junction} is a `Rood' wing,  which has a 3:2 elliptic nose and NACA 0020 tail cylinder joined at the maximum thickness position. This configuration follows that used in the experiment performed by Devenport et al.~\cite{devenport1990}. The Reynolds number based on the airfoil thickness $T$ and the free stream velocity  $U_{\inf}$ is $Re_T = 115000$. Free stream boundary conditions are applied at the far fields, zero gradient boundary condition is applied at the outlet, and non-slip boundary conditions are applied at the walls. The mean flow
is symmetrical about $x$-$z$ plane, and thus only half of the physical domain is simulated in the RANS simulation.

\begin{figure}[htbp]
  \centering
  \includegraphics[width=0.25\textwidth]{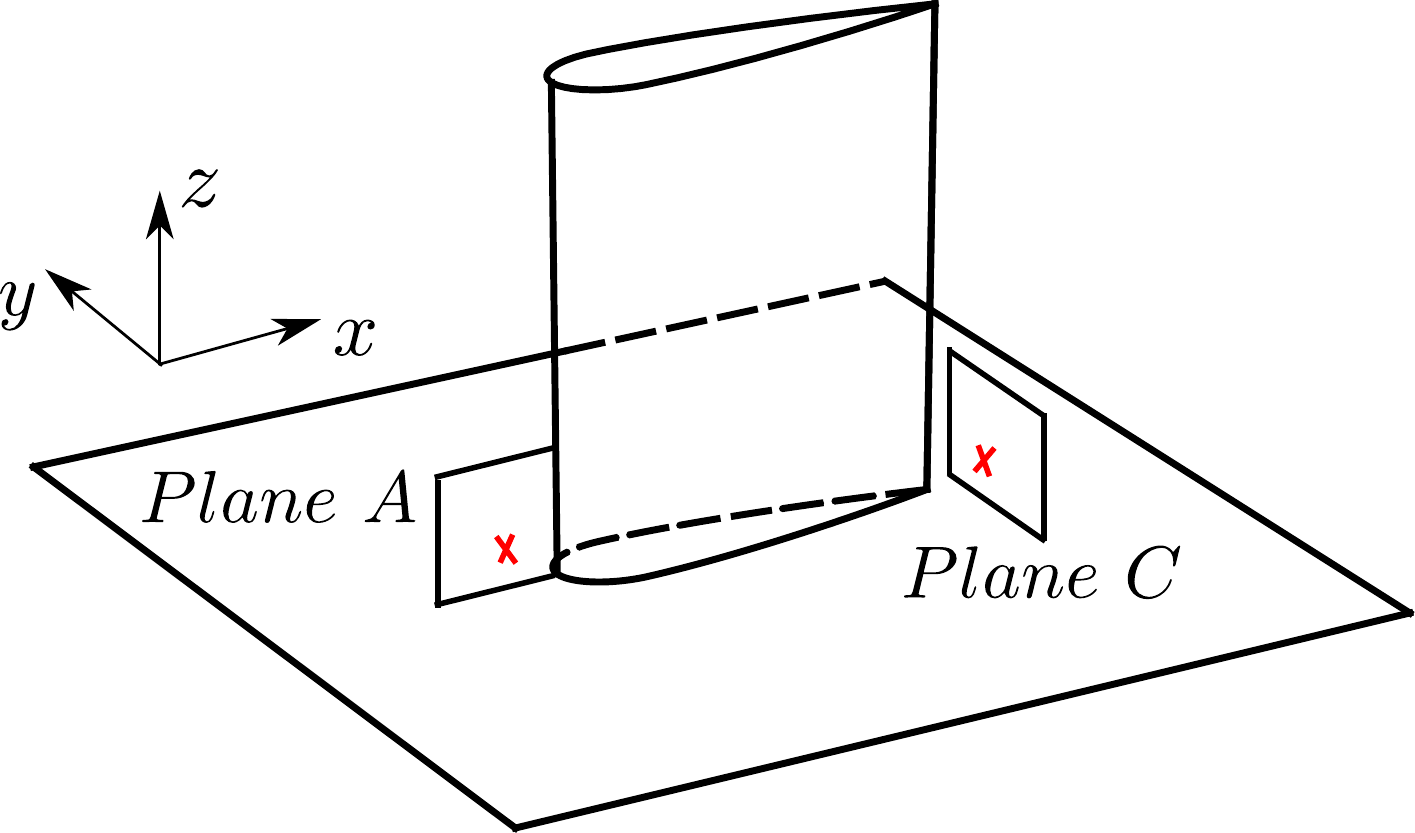}\hspace{5.5em}
  \includegraphics[width=0.25\textwidth]{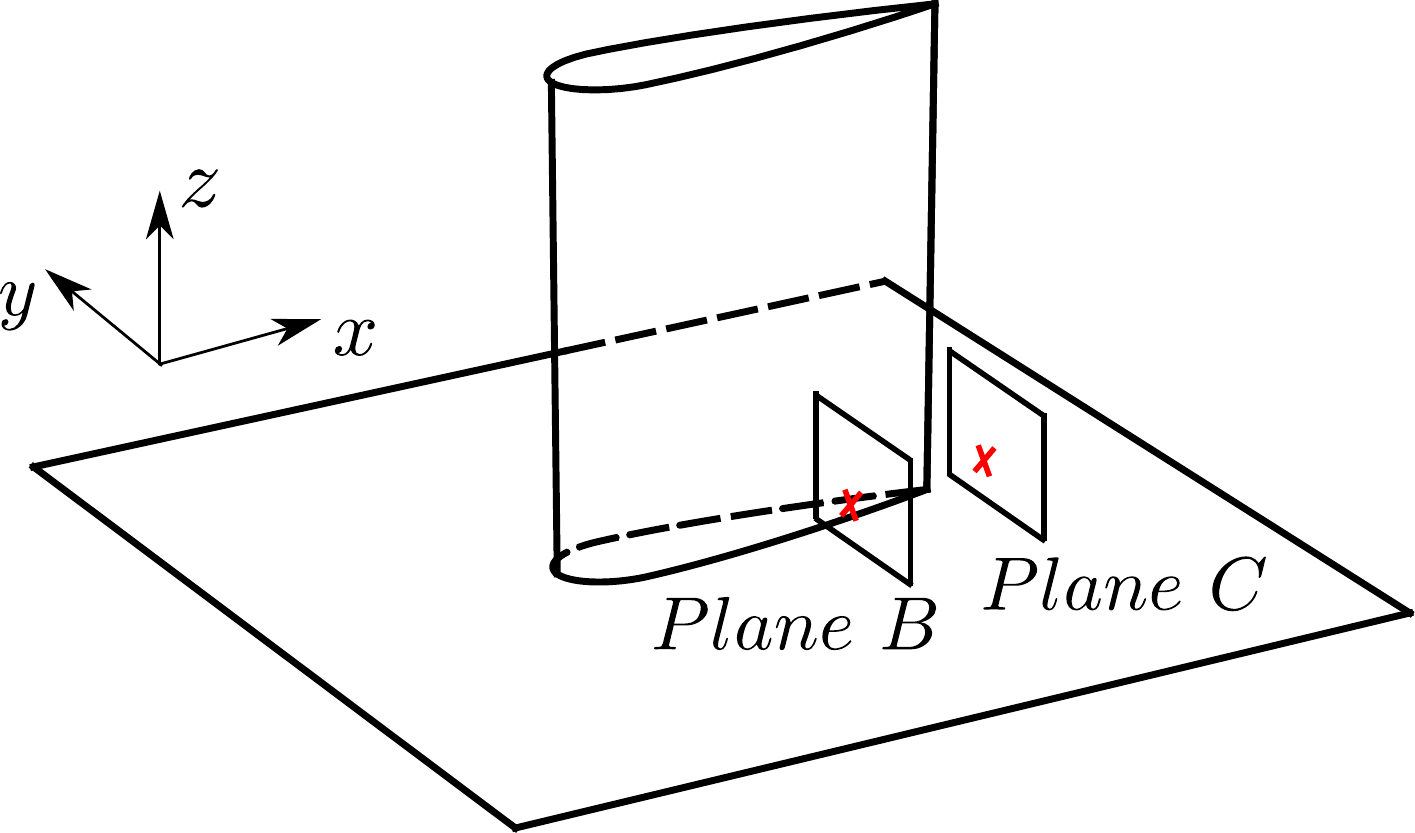}\\
  \subfloat[Correlation between planes A and B]{\includegraphics[width=0.45\textwidth]{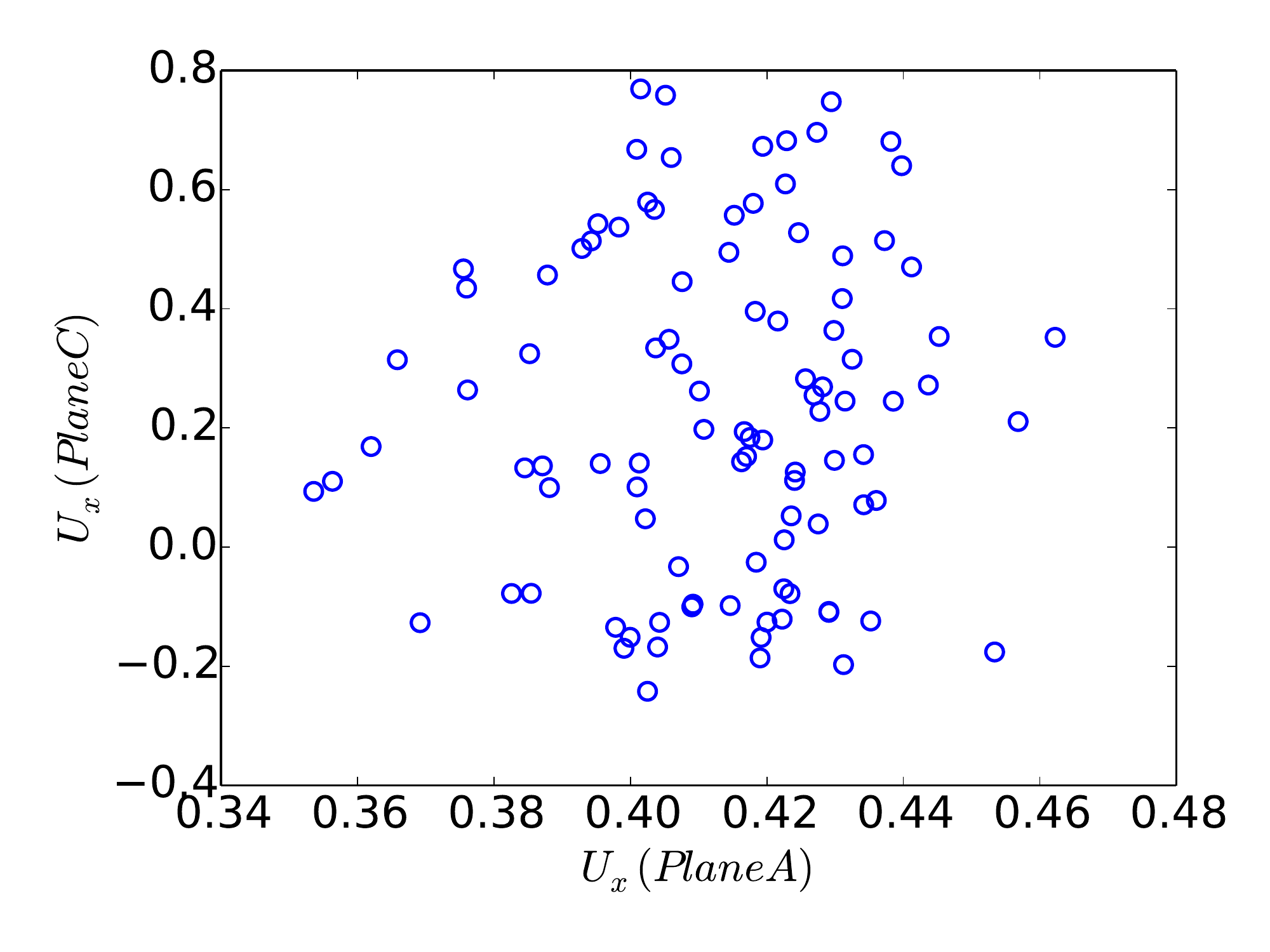}}\hspace{0.5em}
  \subfloat[Correlation between planes B and C]{\includegraphics[width=0.45\textwidth]{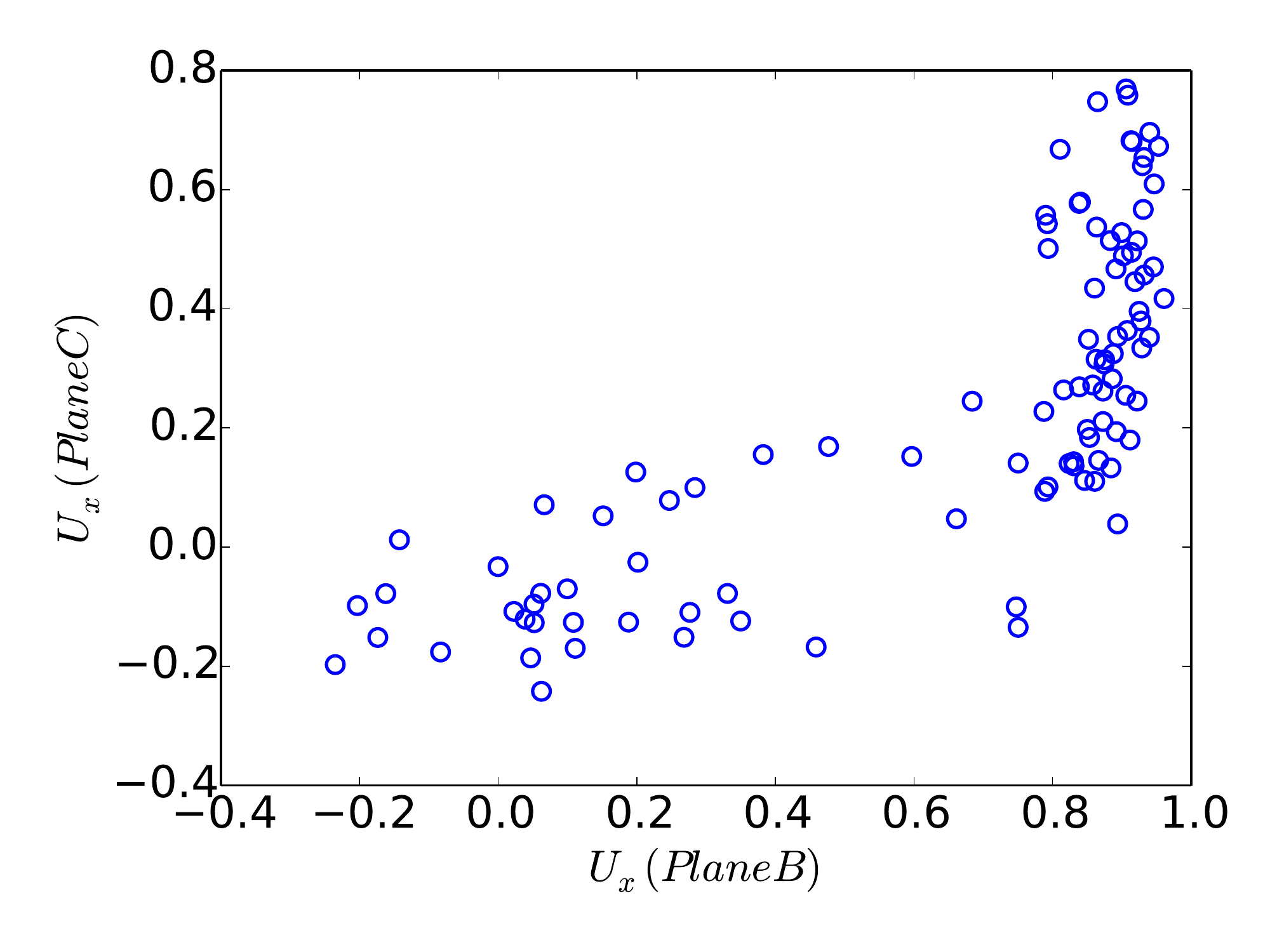}}
  \caption{Mean flow velocity correlation between different planes. Plane A is the symmetrical plane at the upstream of leading edge. Plane B is the secondary flow plane at $x/T=3.187$ within the junction. Plane C is the secondary plane at $x/T=4.4618$, which is at the downstream of trailing edge. The locations of these three planes are chosen based on the available experimental data~\cite{devenport1990}.}
  \label{fig:corr}
\end{figure}

Velocity observations are marked as crosses ($\times$) in Fig.~\ref{fig:domain_junction}, which are generated by adding Gaussian random noises with standard
deviation $\sigma_{obs}$ to the experimental data. According to the experimental results~\cite{devenport1990}, the uncertainties of streamwise velocity $U$ and secondary velocity $W$ are 5\% and 7.2\%, respectively. However, it was also mentioned that some bias error not included in this uncertainty estimation. Therefore, the standard deviation $\sigma_{obs}$ is set as 10\% of the true mean value in this work, which is a total estimation of the uncertainty in the experimental data. The arrangement of observations are based on the analysis of mean flow correlation as shown in Fig.~\ref{fig:corr}. Plane A is the symmetrical plane upstream of leading edge, where the horseshoe vortex system develops. Plane C is the secondary flow plane downstream of the corner region of the trailing edge. Plane B is another secondary flow plane within the junction. It can be seen that the velocity correlation between plane A and plane C is weak, which indicates that the observation information from plane A has little influence upon the mean flow field around the corner separation region. Such weak correlation is also confirmed by a recent NASA experiment~\cite{rumsey2016}. In contrast, the correlation between plane B and plane C is much stronger as shown in Fig.~\ref{fig:corr}b. Therefore, the observation data from plane B would have more influence upon the inference performance in plane C. It should be noted that the Ensemble Kalman Filter used in this framework relies on the mean flow correlation between observation locations and the regions without observations. Since both the horseshoe vortex system and the possible corner separation are of interest in this work, the observations are arranged at both plane A and plane B as shown in Fig.~\ref{fig:domain_junction} to ensure the correlation between the observation locations and the regions of interest. 

\begin{figure}[htbp]
  \centering
    \subfloat[mode 1]{\includegraphics[width=0.32\linewidth]{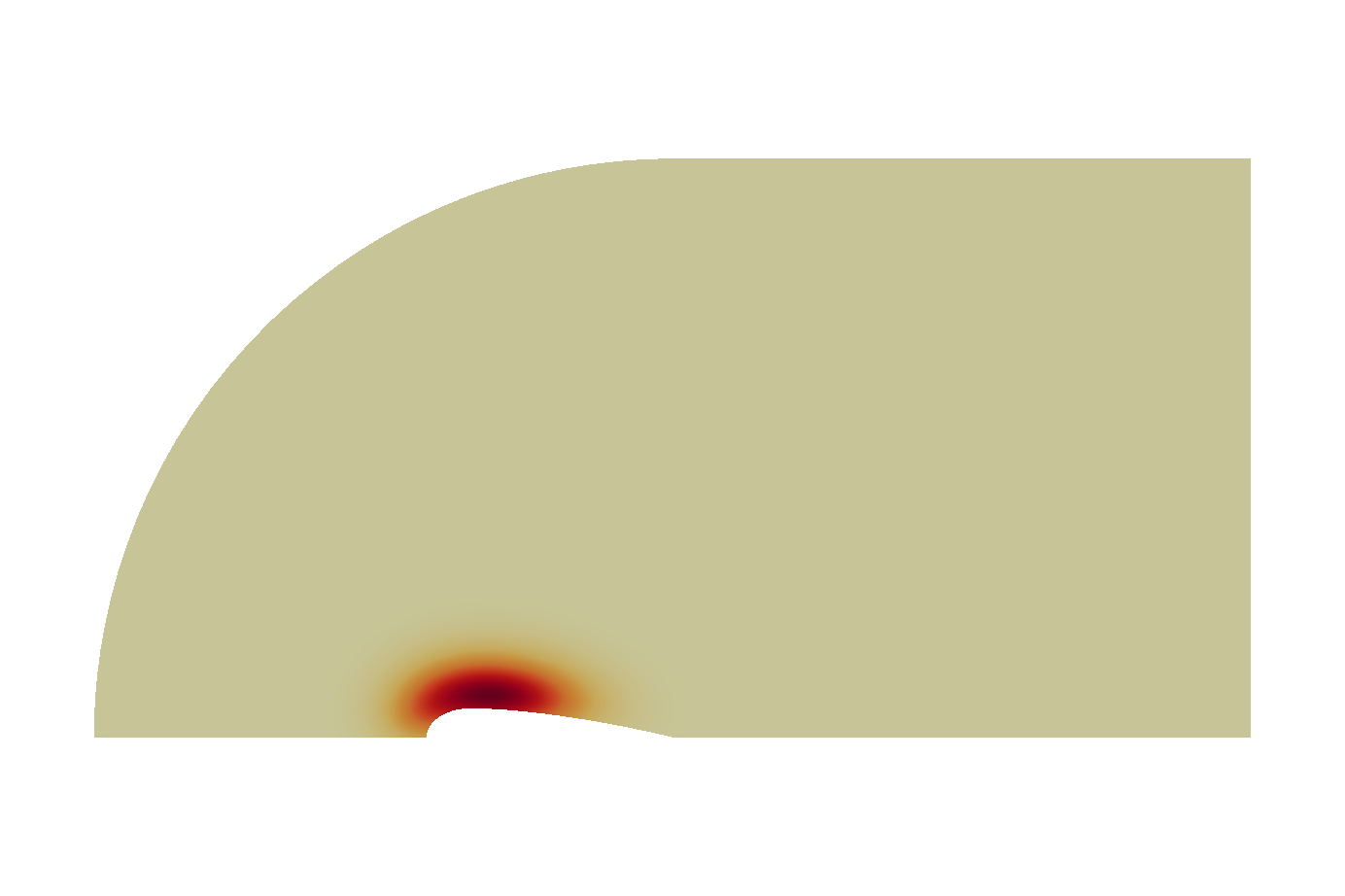}}\hspace{0.1em}
    \subfloat[mode 2]{\includegraphics[width=0.32\linewidth]{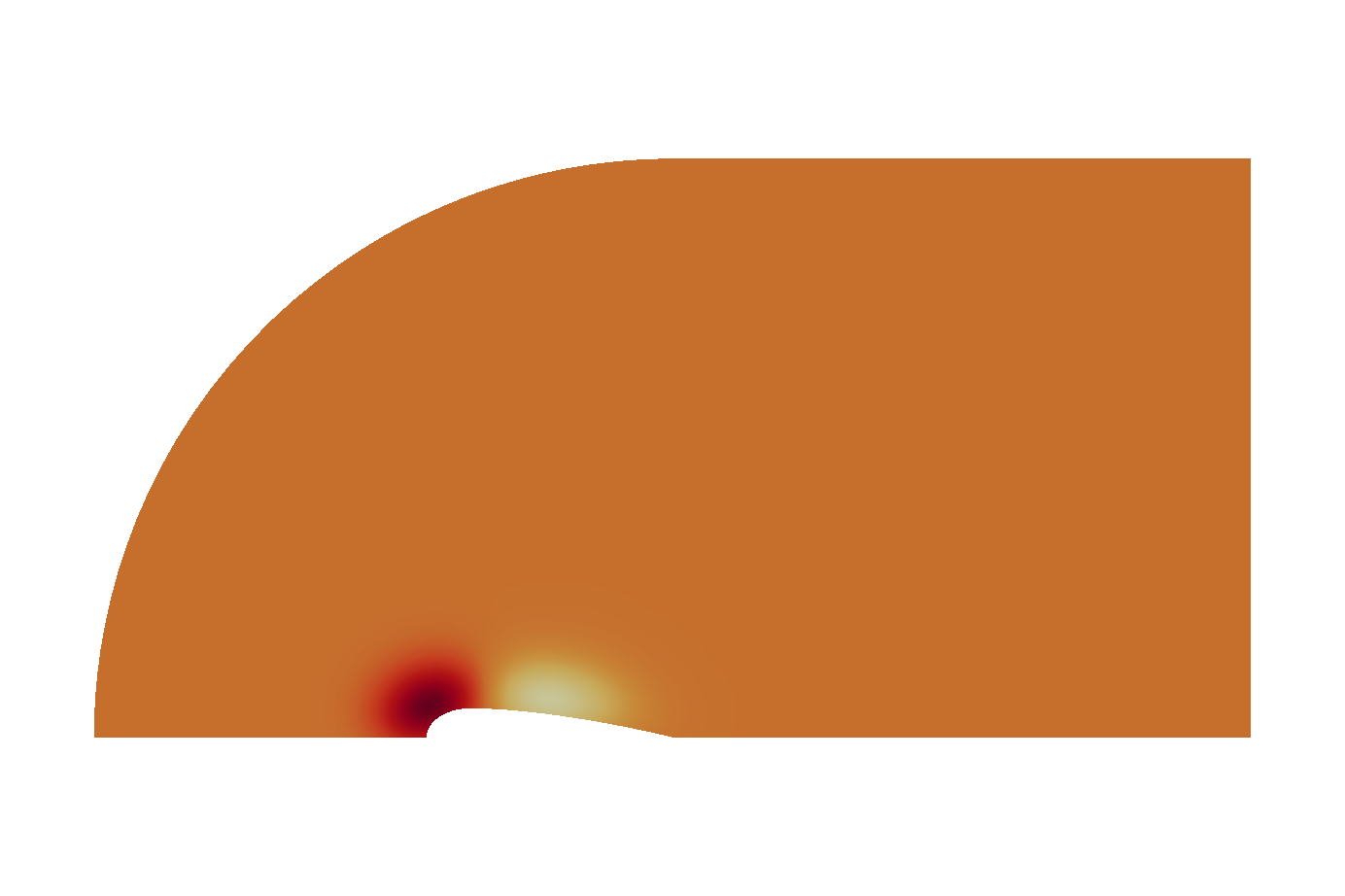}}\hspace{0.1em}
    \subfloat[mode 3]{\includegraphics[width=0.32\linewidth]{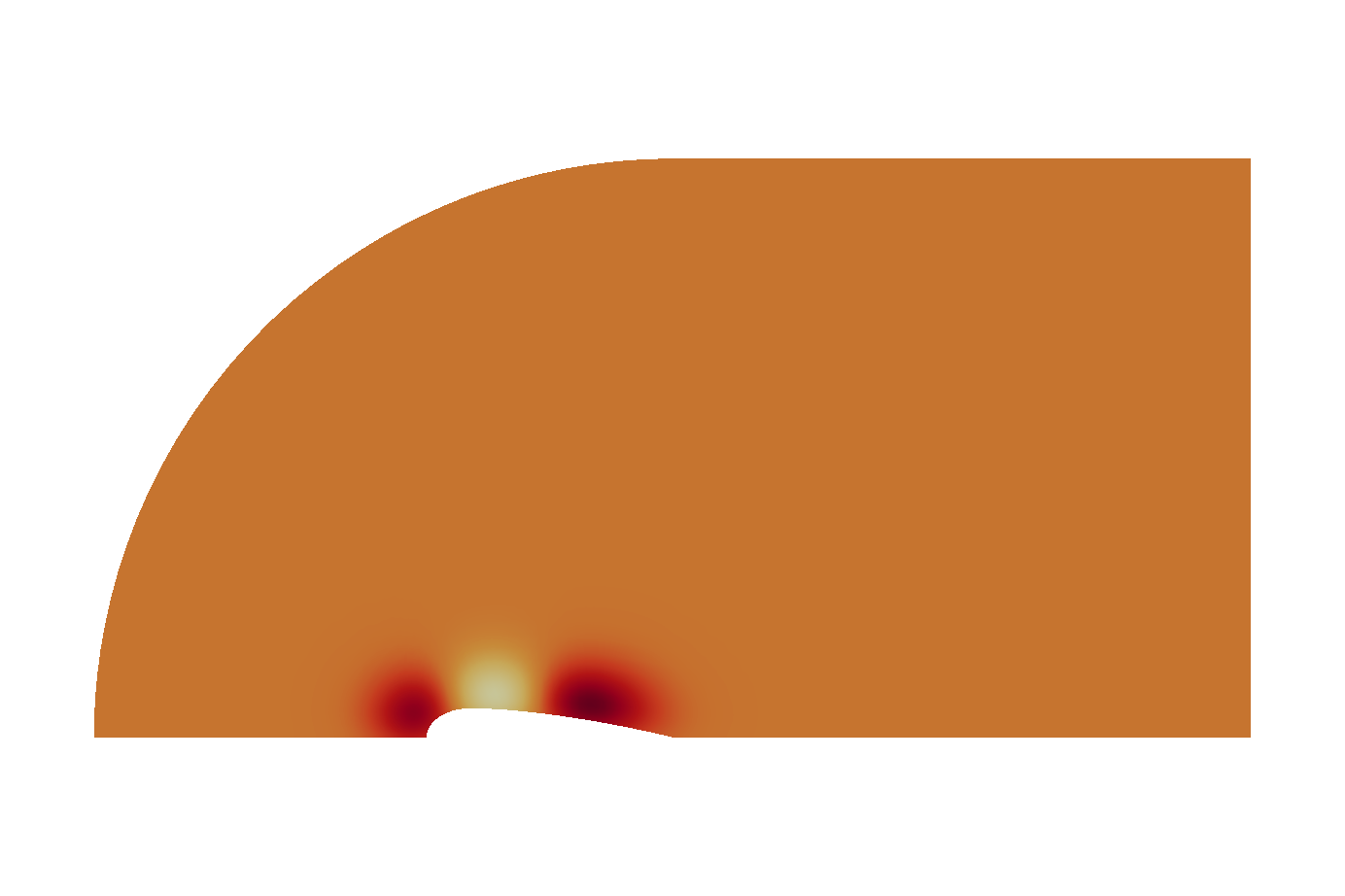}}\hspace{0.1em} \\
    \subfloat[mode 4]{\includegraphics[width=0.32\linewidth]{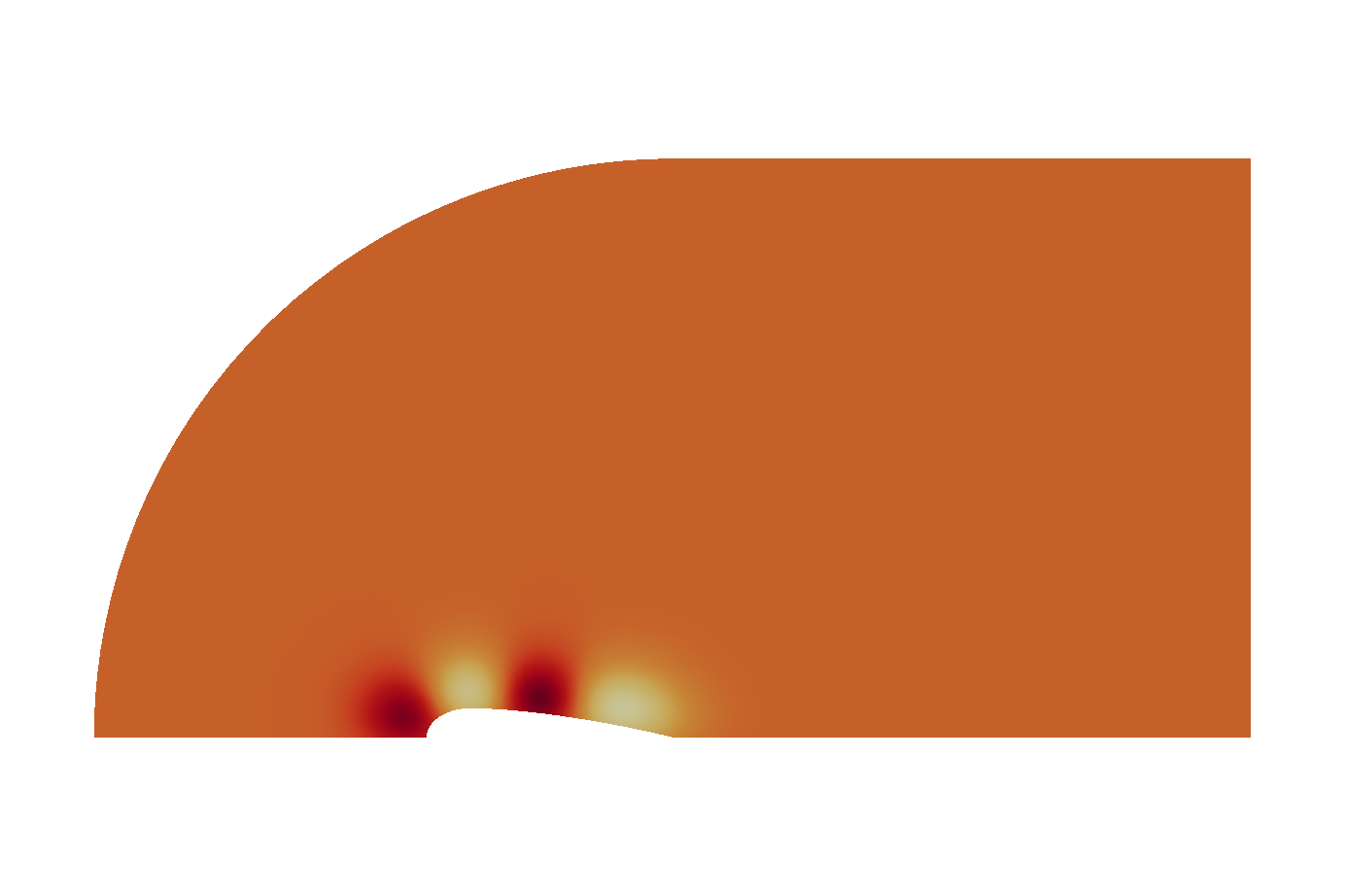}}\hspace{0.1em}
    \subfloat[mode 5]{\includegraphics[width=0.32\linewidth]{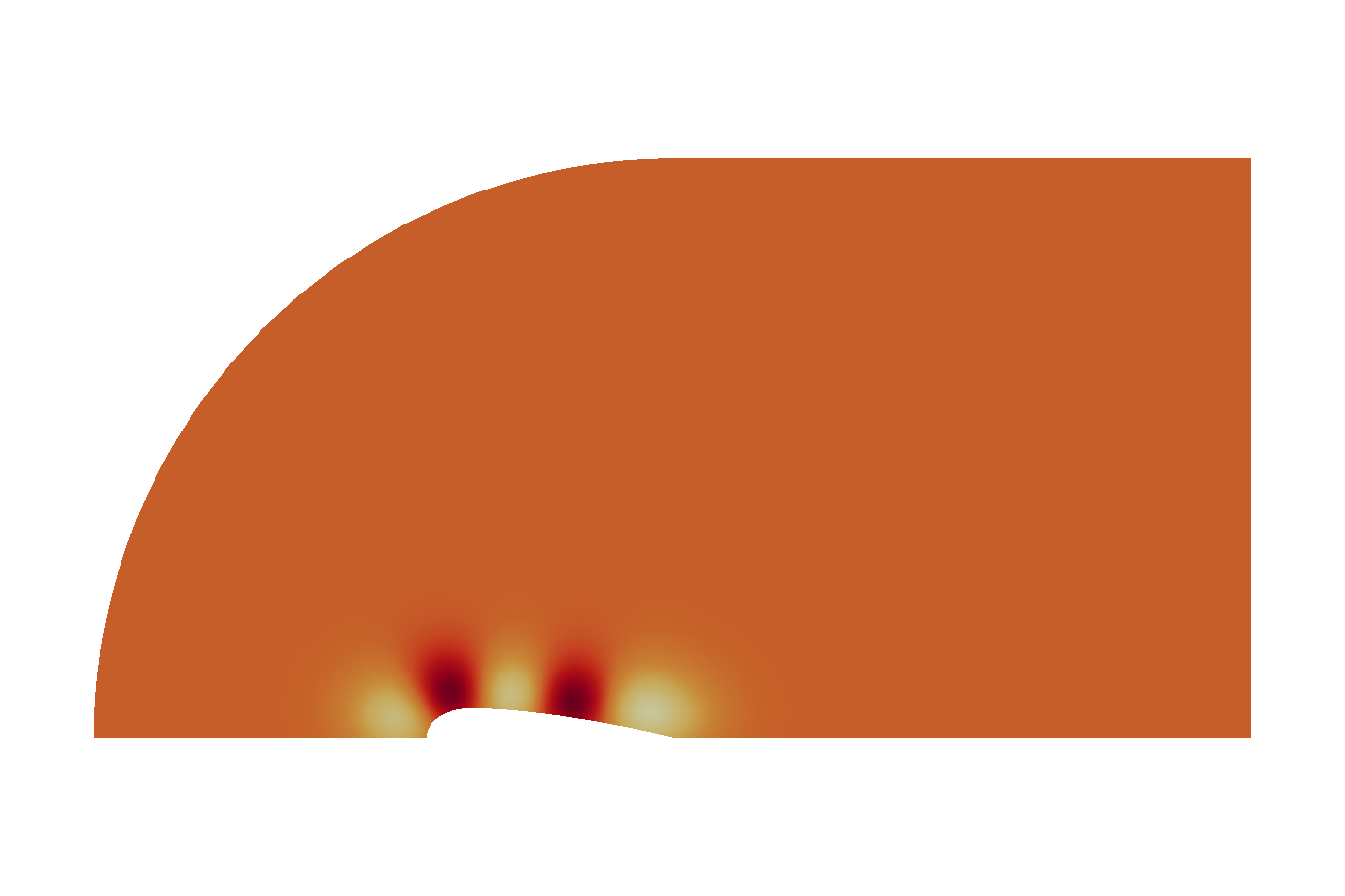}}\hspace{0.1em}
    \subfloat[mode 6]{\includegraphics[width=0.32\linewidth]{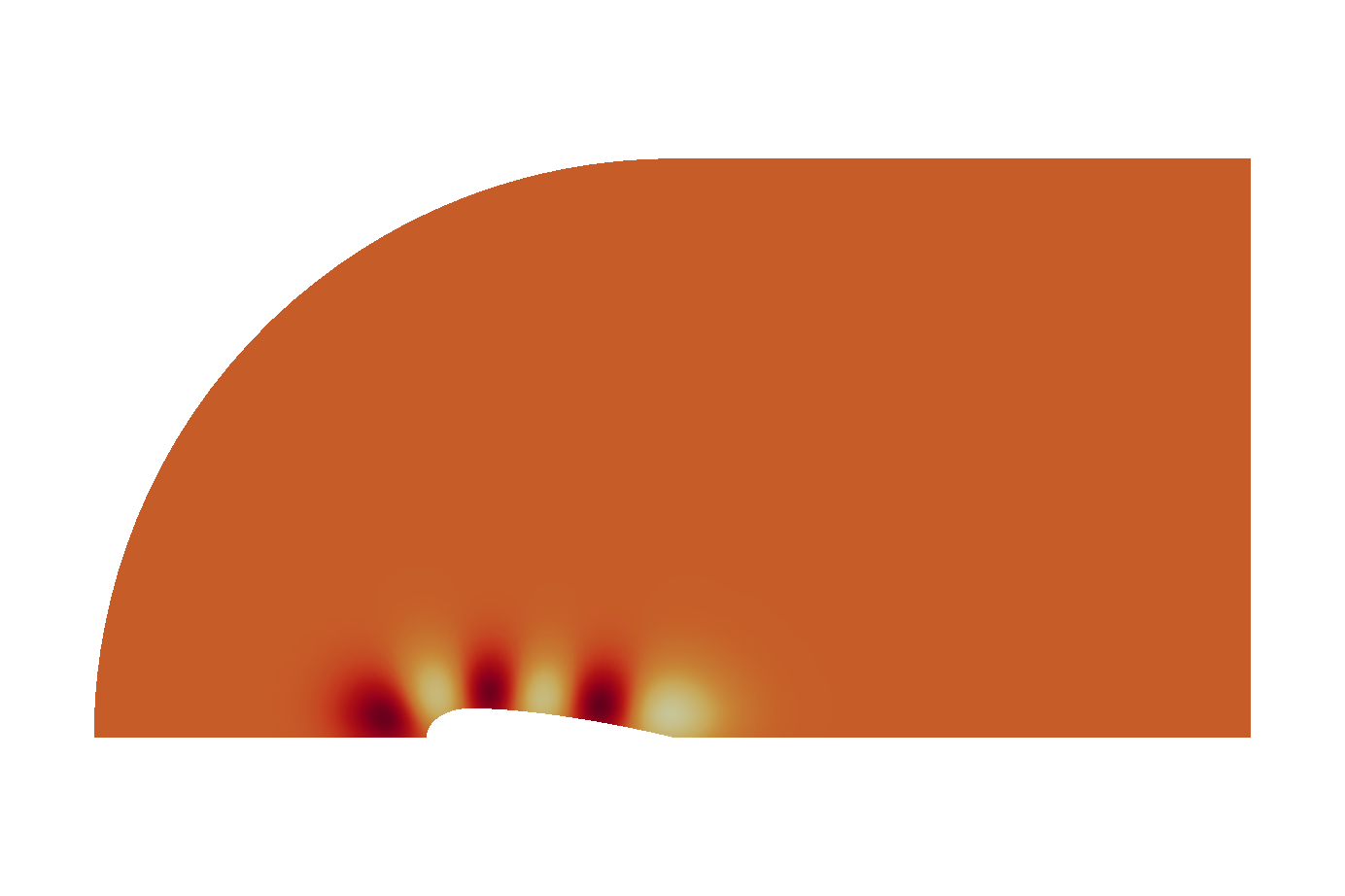}}\hspace{0.1em} \\
    \caption{Illutstration of KL expansion modes of the perturbation field. All the modes have been
      shifted and scaled into the range between 0 (lightest) and 1 (darkest) to facilitate
      presentation, and the legend is thus omitted. Panels (a) to (f) represent modes 1 to 6,
      respectively. Lower modes are more important.}
  \label{fig:modes}
\end{figure}

The perturbations $\delta^\xi$, $\delta^\eta$ and $\delta^k$ are modeled as non-stationary Gaussian process. The variance field $\sigma(x)$ of the Gaussian process is based on physical prior knowledge, i.e., the RANS prediction is more unreliable at the near wall region of the airfoil. To achieve a more compact representation of the Gaussian process, Karhunen-Loève (KL) expansion is used in the framework to reconstruct the Gaussian process. Specifically, 6 modes are used in this work. All the modes are illustrated in Fig.~\ref{fig:modes}. It should be noted that these modes do not have any variation along the $y$ direction. This is a simplification in this framework to reduce the number of modes. Although the true Reynolds stress discrepancy field is more likely to be 3D in this flow, the number of modes will grow rapidly if 3D mode is used. Since the amount of observation points and the computational cost are both related to the number of modes, it is impractical to introduce a large amount of modes in real applications. Therefore, we make an assumption that the variation of the discrepancy field around the airfoil is much more complex than that within the boundary layer along the $y$ direction. Based on this assumption, the choice of 2D modes as shown in Fig.~\ref{fig:modes} can be justified.

\subsection{Results at Corner Region}
\label{sec:result-corner}
\begin{figure}[!htbp]
  \centering 
  \includegraphics[width=0.6\textwidth]{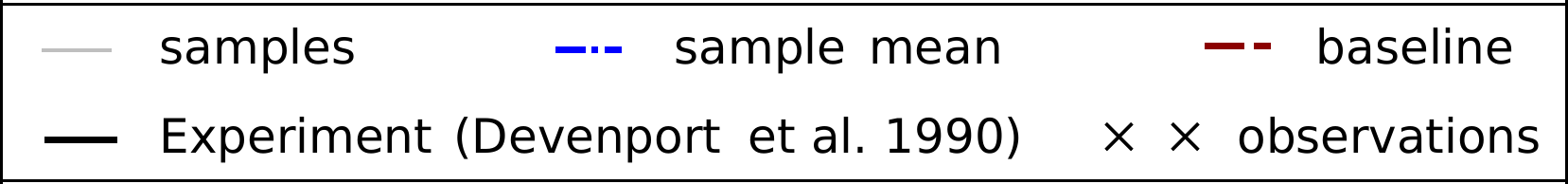}\\
  \vspace{1em}
  \includegraphics[width=0.35\textwidth]{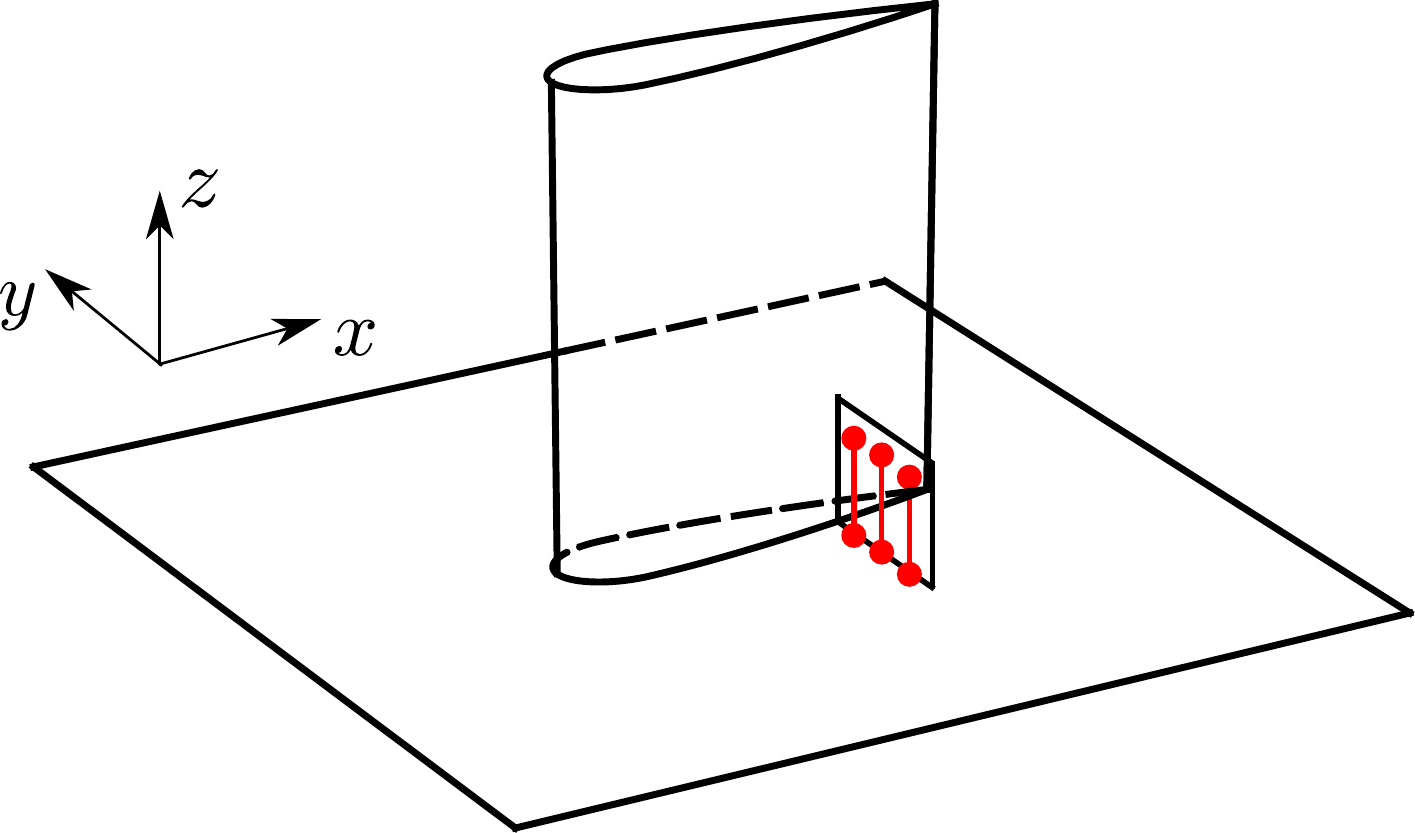}\\
  \subfloat[Secondary velocity $U_y$ at plane $x/T=3.187$]{\includegraphics[width=0.45\textwidth]{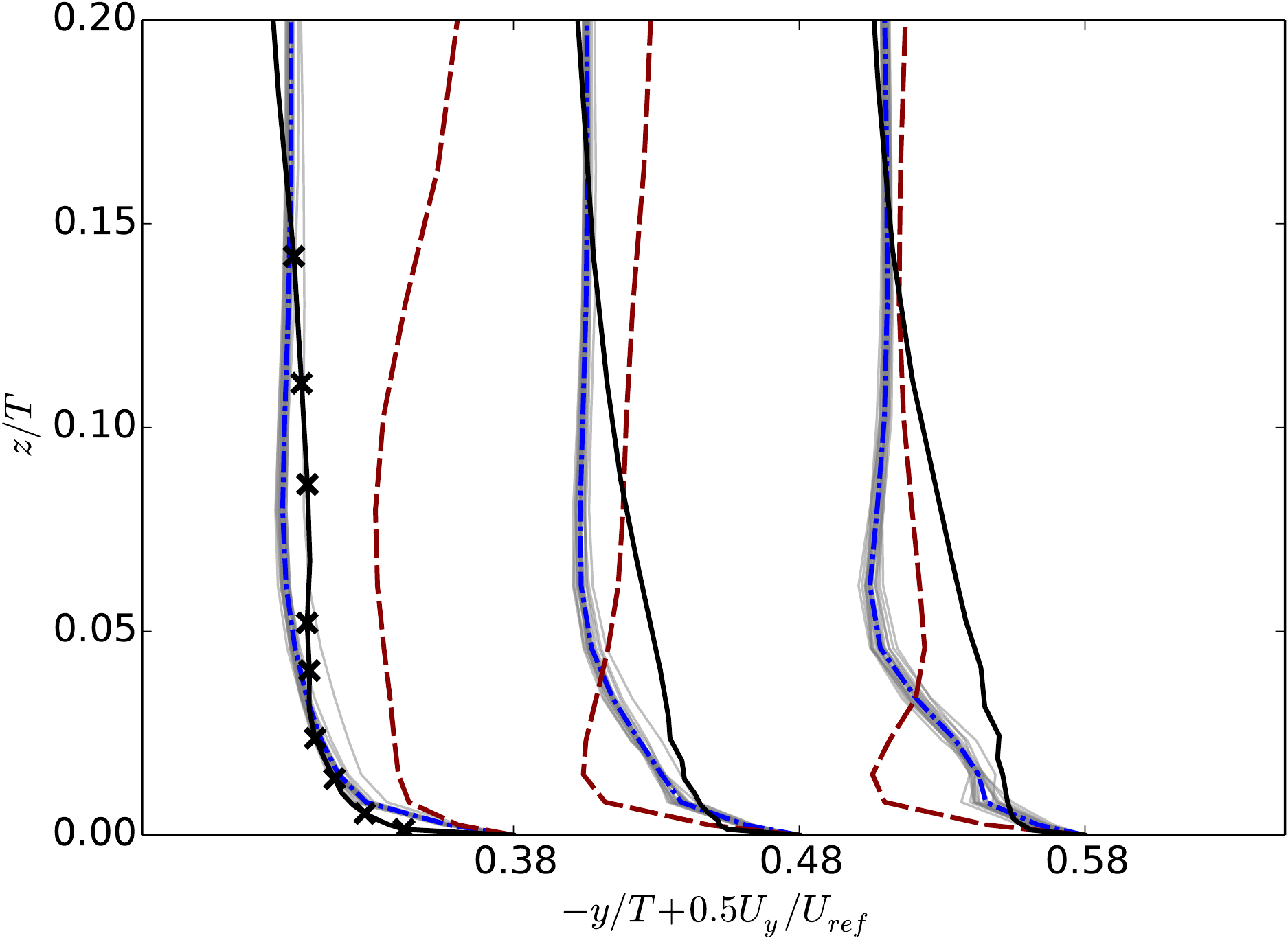}}\hspace{0.5em}
  \subfloat[Secondary velocity $U_z$ at plane $x/T=3.187$]{\includegraphics[width=0.45\textwidth]{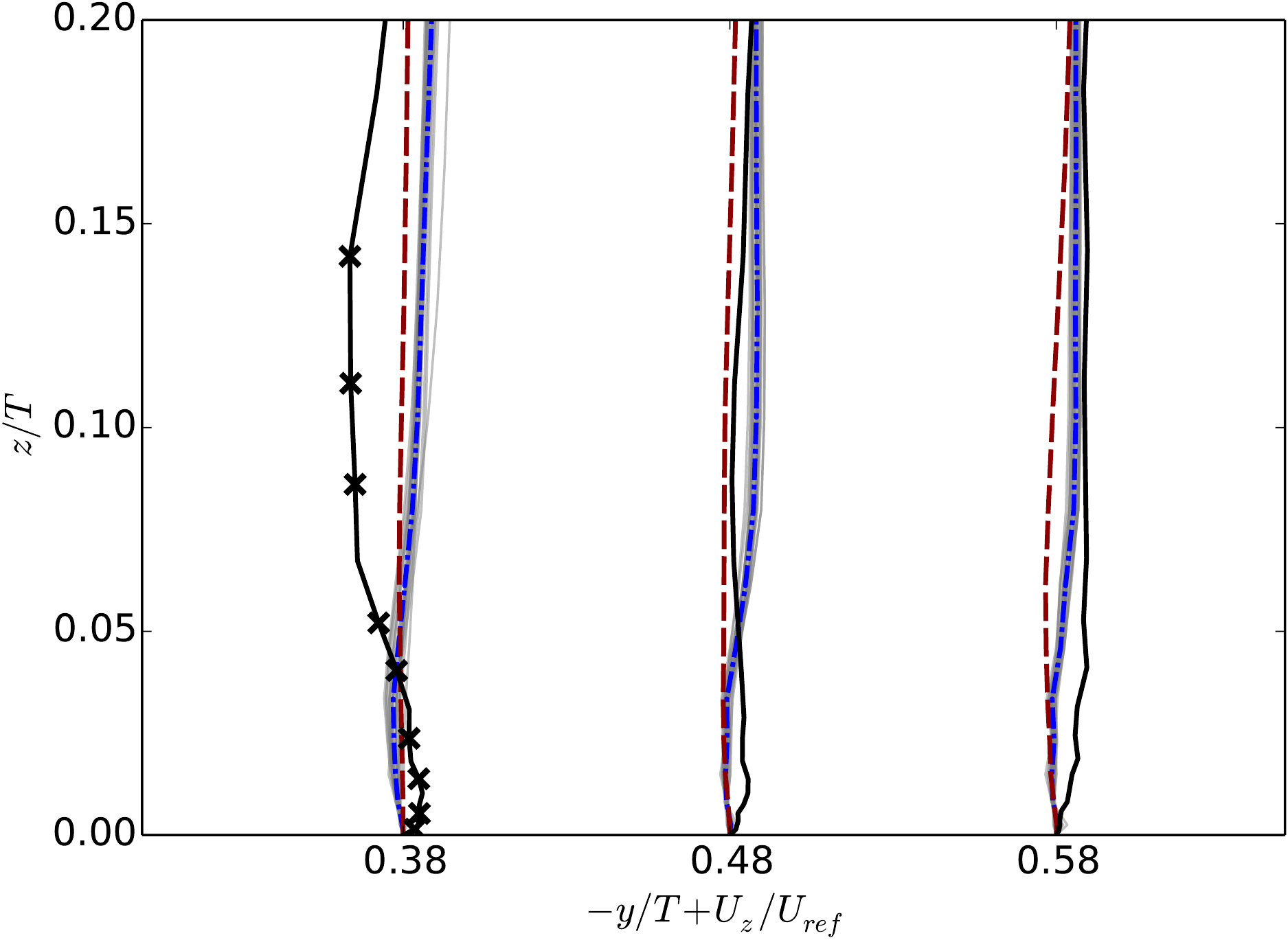}}\
  \caption{Posterior ensemble of secondary velocity at plane B ($x/T=3.187$) along three locations $-y/T=0.38$, $0.48$ and $0.58$ with increasing distance from the airfoil surface. A smaller value of $-y/T$ represents a location closer to the airfoil. The velocity profiles of $U_y$ is scaled by a factor of 0.5 for clarity. Black crosses ($\times$) denote locations where velocity observations are available.}
\label{fig:plane-8}
\end{figure}

The posterior secondary velocity profiles at the plane B ($x/T=3.187$) within the junction are shown in Fig.~\ref{fig:plane-8}. It can be seen that the posterior velocity profile $U_y$ has a much better agreement with the experimental data at the location $-y/T=0.38$, where observation data is available. In addition to the $U_y$ profiles at $-y/T=0.38$ where observation data are available, the posterior velocity profiles $U_y$ at $-y/T=0.48$ and $-y/T=0.58$ are also improved, while over-correction is noticeable. It indicates that the length scale we used to construct the gaussian random field is larger than the true length scale of the mean flow. As a consequence, the mean flow correlation is over-estimated and it leads to some over-correction in the region where observation data are not available. Compared to the posterior velocity $U_y$, the posterior velocity $U_z$ profiles have less improvement as shown in Fig.~\ref{fig:plane-8}b. A possible reason is that the discrepancy field is 2D according to the KL modes as shown in Fig.~\ref{fig:modes}. The complexity of 2D discrepancy field would not exactly satisfy the true discrepancy of Reynolds stress. Therefore, the posterior velocity profiles may not exactly agree with the observation data, which explains the mismatch of $U_z$ at $-y/T=0.38$ where observation data is available.

\begin{figure}[!htbp]
  \centering 
  \includegraphics[width=0.6\textwidth]{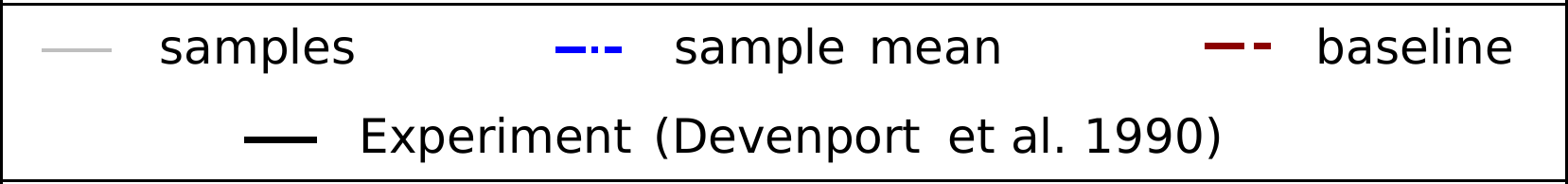}\\
  \vspace{1em}
  \includegraphics[width=0.35\textwidth]{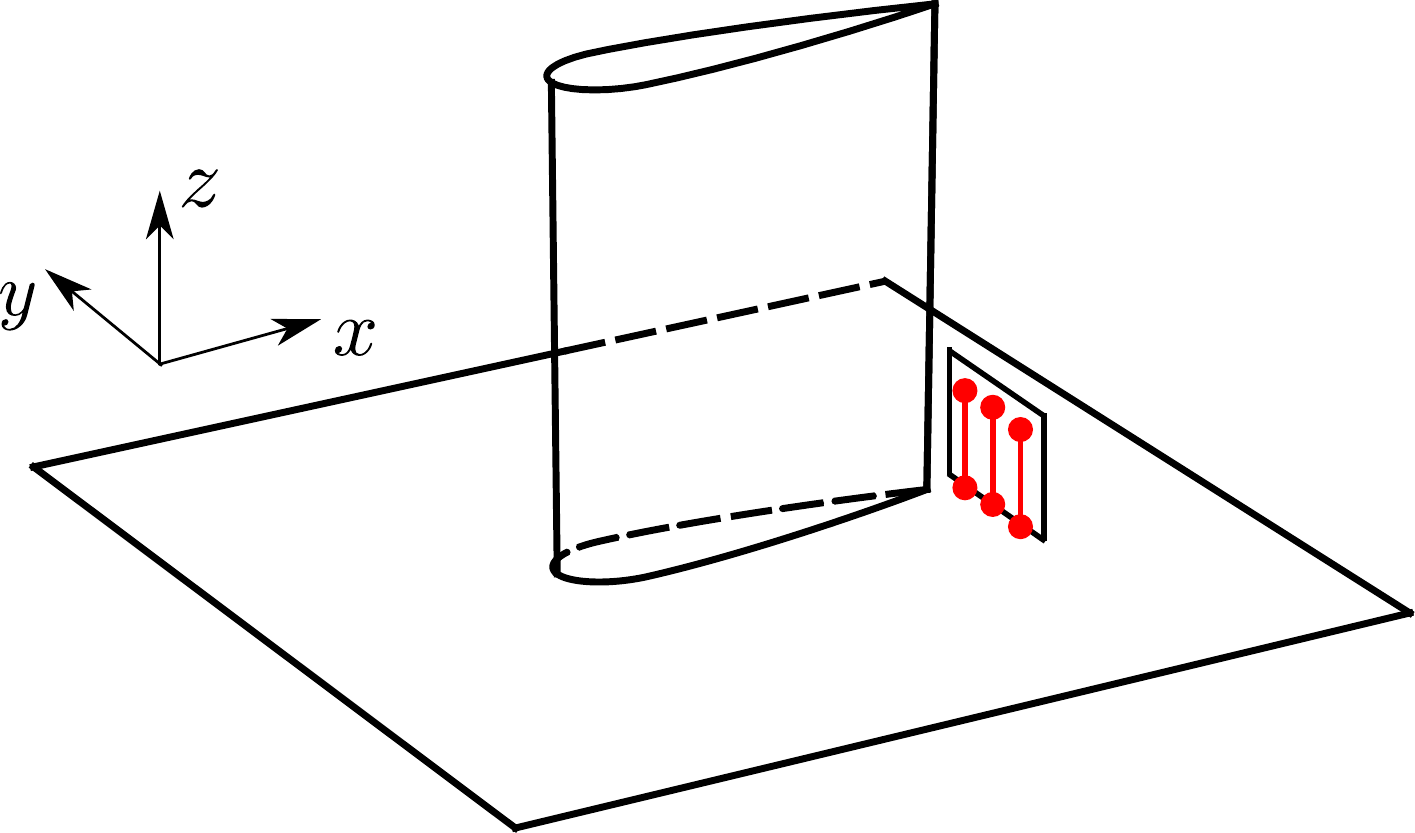}\\
  \subfloat[Secondary velocity $U_y$ at wake $x/T=4.4618$]{\includegraphics[width=0.45\textwidth]{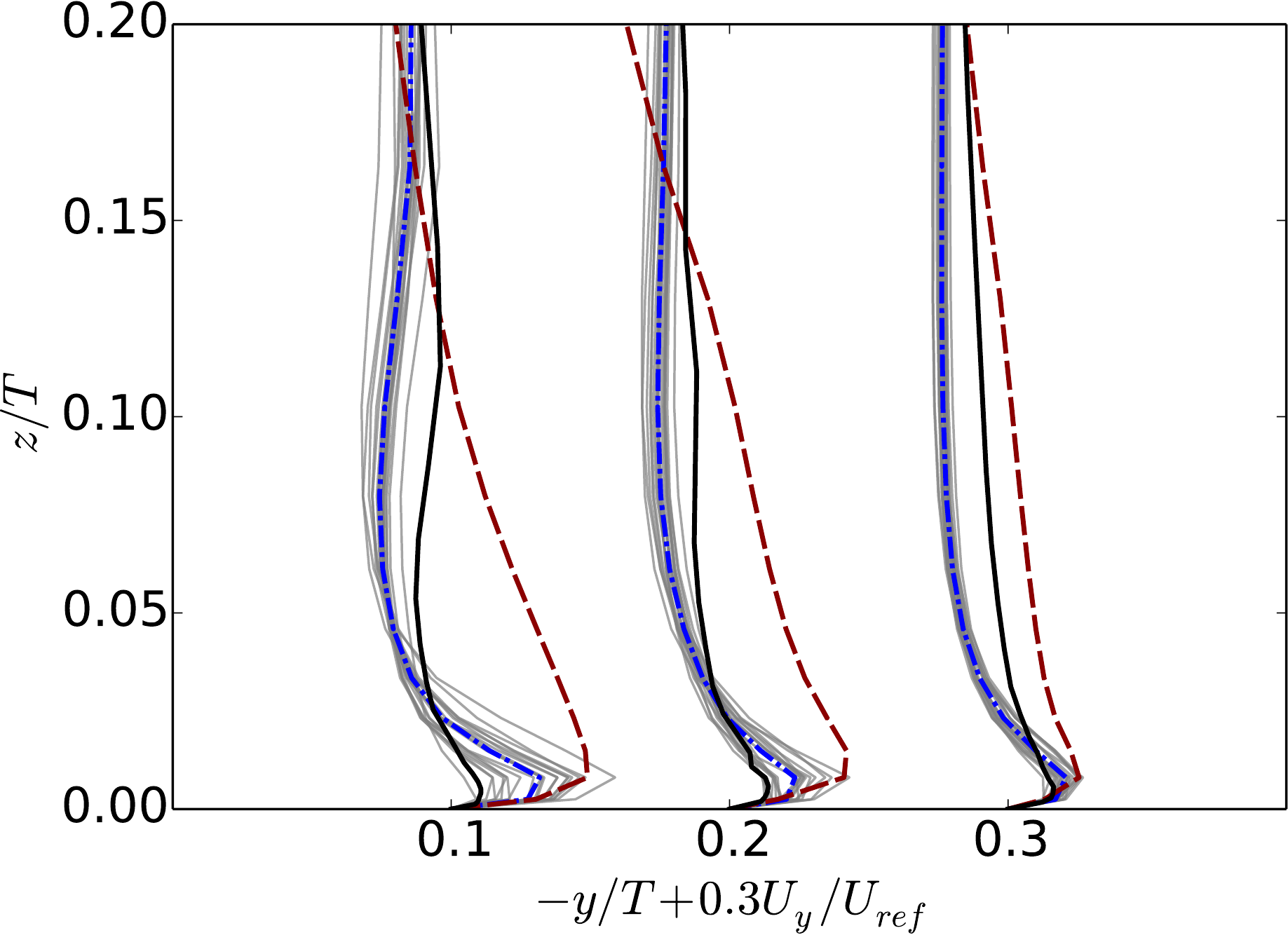}}\hspace{0.5em}
  \subfloat[Secondary velocity $U_z$ at wake $x/T=4.4618$]{\includegraphics[width=0.45\textwidth]{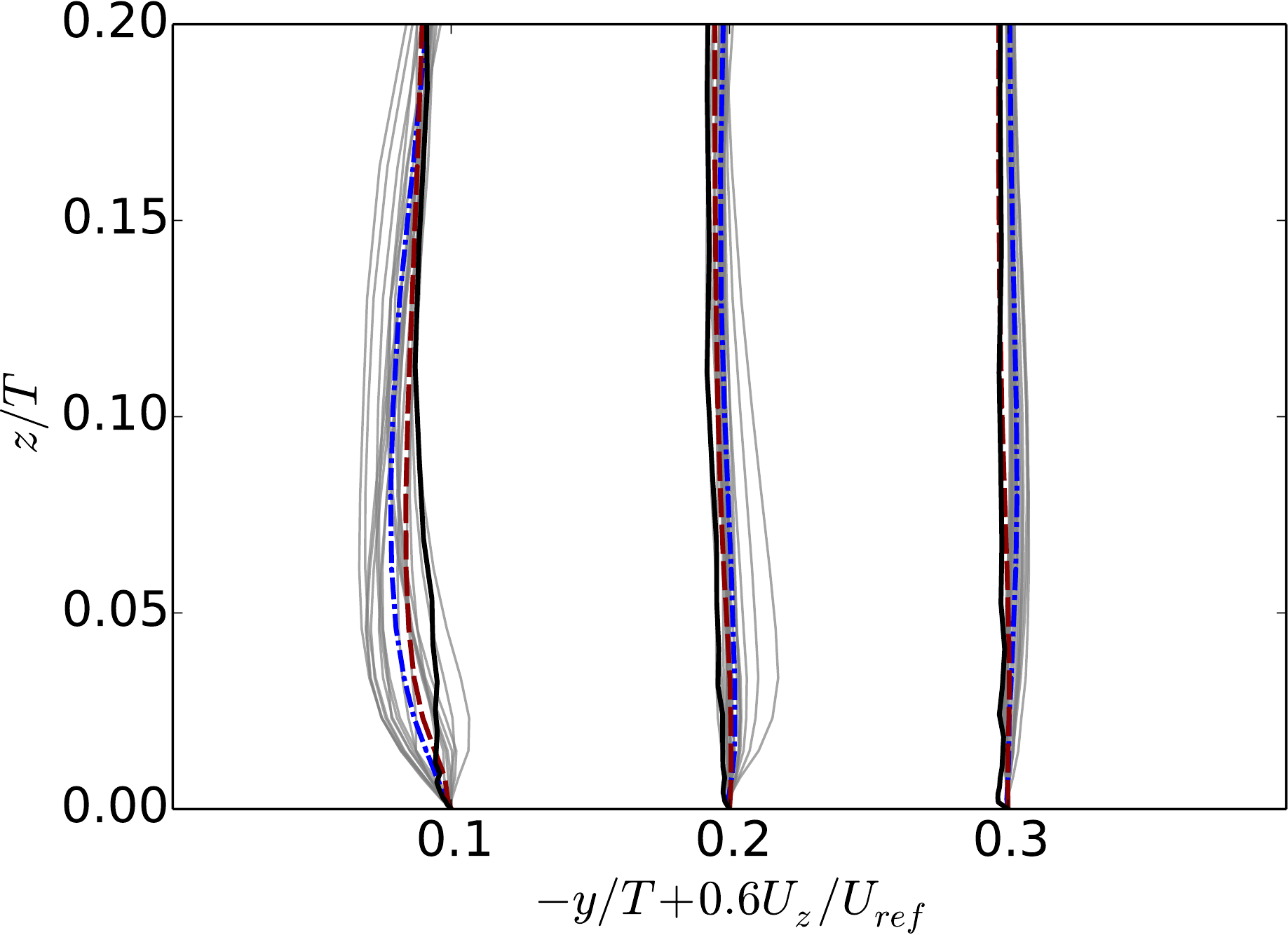}}\
  \caption{Posterior ensemble of secondary velocity at plane C ($x/T=4.4618$) along three locations $-y/T=0.38$, $0.48$ and $0.58$, with increasing distance from the airfoil surface. A smaller value of $-y/T$ represents a location closer to the airfoil. The velocity profiles are scaled by a factor of 0.5 for clarity. No velocity observation is applied at this plane.}
\label{fig:plane-10}
\end{figure}

Figure~\ref{fig:plane-10} shows the posterior secondary velocity profiles at plane C in the wake downstream of the trailing edge. Compared to the baseline RANS prediction, the posterior velocity profiles show significant improvement, even though no observation data is available at this plane. Such improvement is due to the correlation between this plane and plane B where observation data is available. Specifically, the prediction of posterior velocity profiles at upstream plane B shows better agreement with experimental data in Fig.~\ref{fig:plane-8}, especially for the $U_y$ profiles at the region near airfoil ($-y/T=0.38$). It demonstrates that the Bayesian framework can provide better prediction for this complex flow problem, even though there is no local observation data available. 

\begin{figure}[!htbp]
  \centering 
  \includegraphics[width=0.43\textwidth]{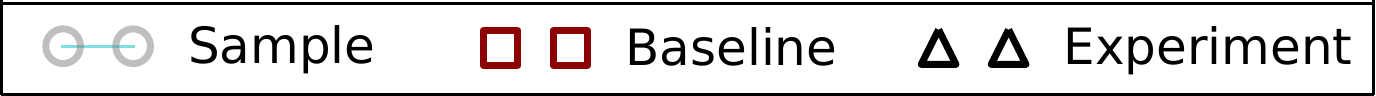}\\
  \vspace{1em}
  \includegraphics[width=0.35\textwidth]{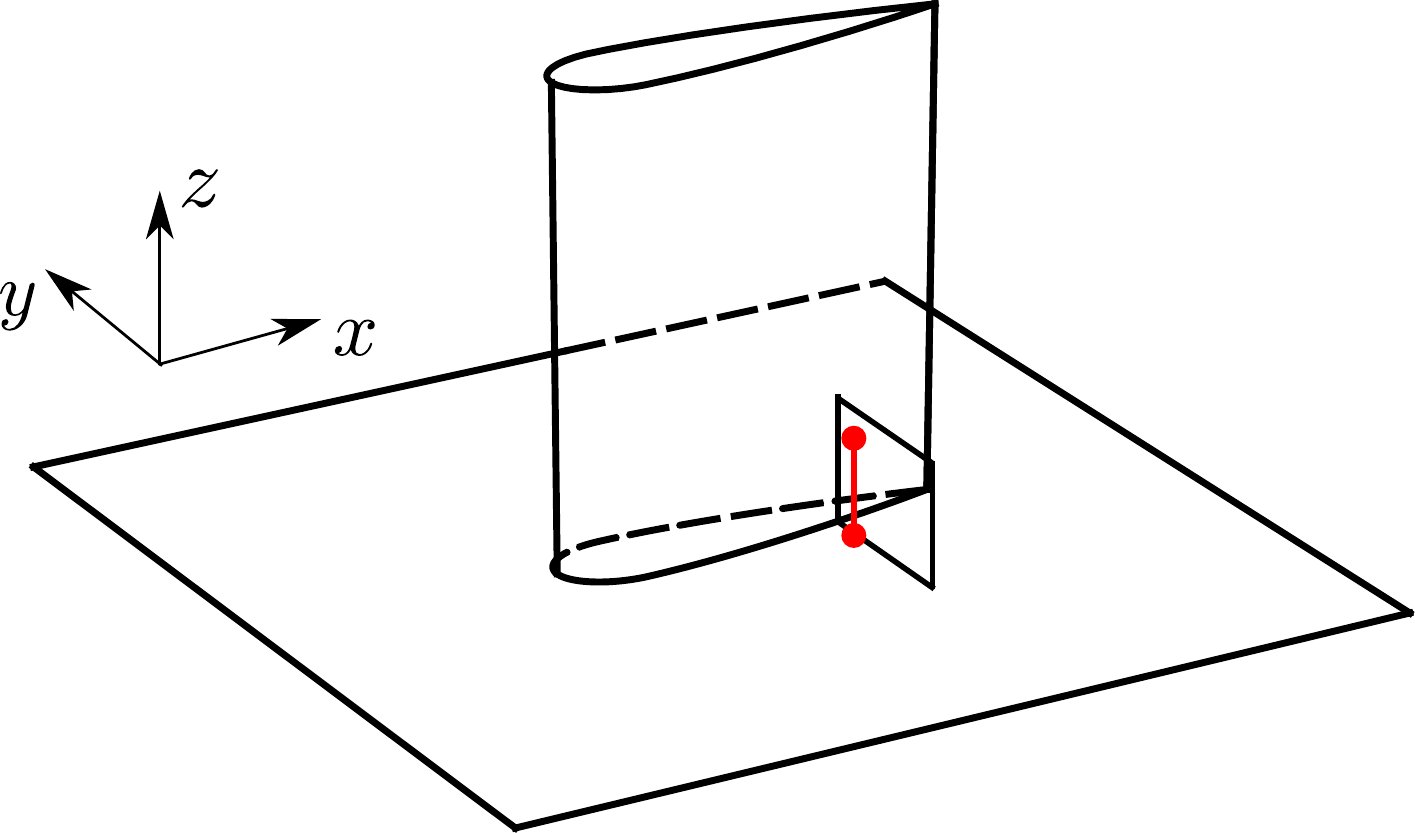}\\
  \subfloat[Prior]{\includegraphics[width=0.45\textwidth]{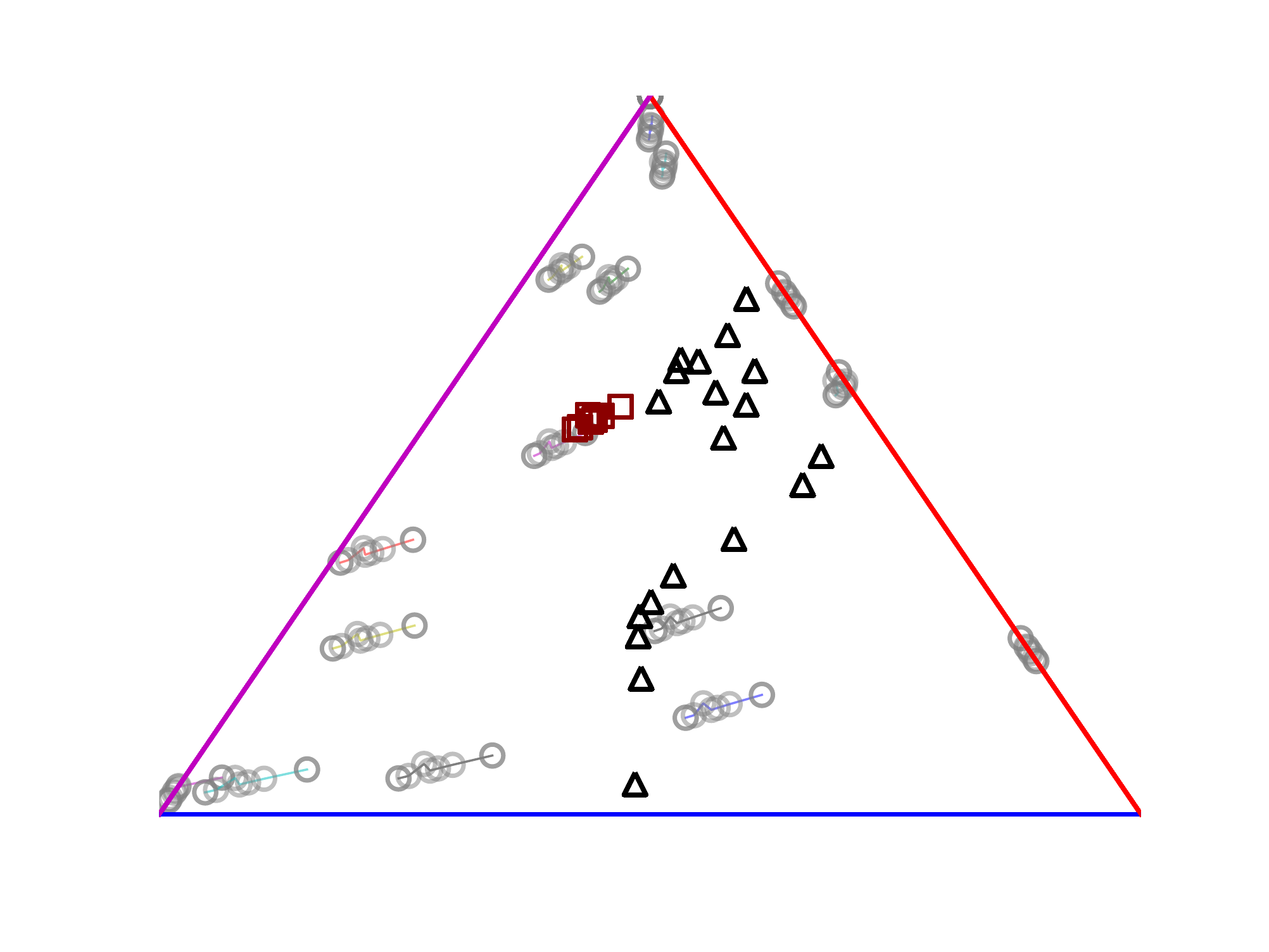}}\hspace{0.5em}
  \subfloat[Posterior]{\includegraphics[width=0.45\textwidth]{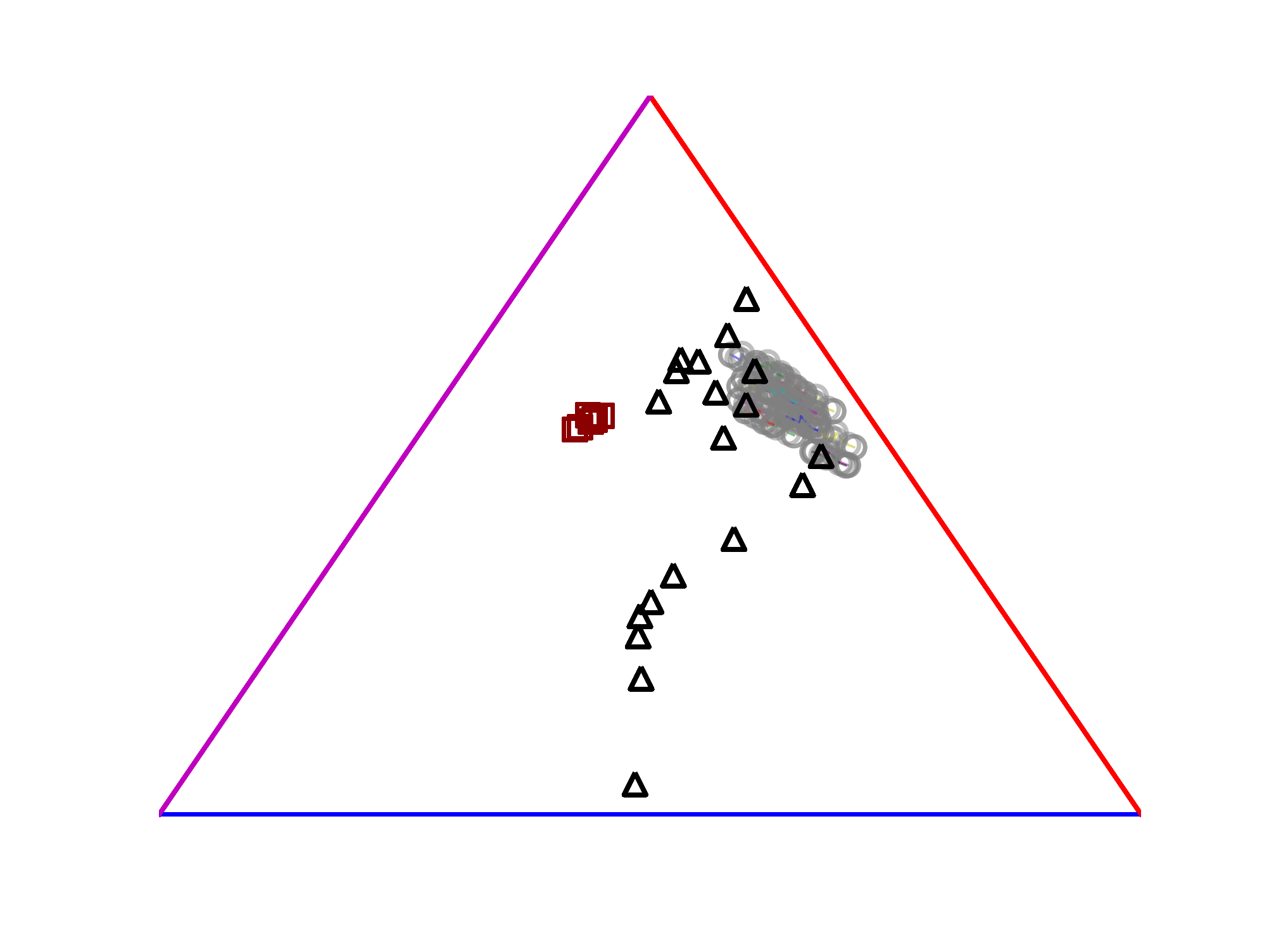}}\
  \caption{Prior and posterior ensemble of Reynolds stress in Barycentric triangle. The Reynolds stresses are sampled at plane B ($x/T=3.187$) along $-y/T=0.38$. The Reynolds stresses from the same sample are linked by line to illustrate the distribution of different samples.}
\label{fig:tau_tri_plane8}
\end{figure}

The comparison of prior and posterior Reynolds stress is shown in Fig.~\ref{fig:tau_tri_plane8} in Barycentric triangle. The Reynolds stress are sampled from the line along $-y/T=0.38$ at plane B ($x/T=3.1817$). The length along $z$ direction of the sample region is about 0.2 thickness of the airfoil. It can be seen from Fig.~\ref{fig:tau_tri_plane8} that the baseline RANS Reynolds stress is quite different from the experimental data. By injecting uncertainties into the baseline RANS Reynolds stresses, the prior ensemble of Reynolds stress explores most part of Barycentric triangle as shown in Fig.~\ref{fig:tau_tri_plane8}a. Compared to the prior Reynolds stresses, the posterior ones have a better agreement with the experimental data, as shown in Fig.~\ref{fig:tau_tri_plane8}b. However, it should be noted that the posterior Reynolds stresses still do not cover the experimental data. A possible reason is that the baseline RANS Reynolds stresses have a more clustered distribution than the experimental data in Barycentric triangle. Such clustered distribution is largely preserved due to the choice of length scale in constructing the Gaussian random field for the perturbation. Specifically, the length scale of Gaussian random field is chosen as the thickness of the airfoil, which accounts for the estimation of mean flow correlation when KL expansion is performed. On the other hand, it can be seen in Fig.~\ref{fig:tau_tri_plane8} that the discrepancy of Reynolds stress anisotropy still has large variation within the sampled region. Such variation is not considered in this work for two reasons. First, the variation at such small length scale indicates much more modes of KL expansion, which increase the degree of freedom of the unknown parameters and require much more observation data to constrain those unknown parameters. However, a large amount of observation data is usually impractical in most engineering applications. Second, the variation at this small region is difficult to estimate beforehand without a comprehensive experimental database. Therefore, it is not feasible to take such variation into account for most engineering applications of interest. Due to these two reasons, the variation of Reynolds stress discrepancy at such small scale is not considered in this work when the uncertainty is injected into the Reynolds stress. Consequently, the relative locations of Reynolds stresses is largely preserved in the Barycentric triangle if the physical distance is small. Such limitation is referred to as the preservation of relative locations of Reynolds stress in the rest part of this work. It indicates that the perturbed Reynolds stress will not explore the true uncertainty space as shown in Fig.~\ref{fig:tau_tri_plane8}, and it is expected that the inferred Reynolds stress would not exactly match with the experimental data in the Barycentric triangle. Although it is a limitation of this framework, it should be noted that the posterior Reynolds stress anisotropy indeed demonstrates some improvement as indicated by the locations in Barycentric triangle in Fig.~\ref{fig:tau_tri_plane8}b. In addition, the mean velocity profiles are also improved as shown in Fig.~\ref{fig:plane-8} and Fig.~\ref{fig:plane-10}, which is of more interest in many engineering applications.

\begin{figure}[!htbp]
  \centering 
  \includegraphics[width=0.6\textwidth]{U-legend-noObs}\\
  \vspace{1em}
  \includegraphics[width=0.35\textwidth]{rood-3D-Plane8}\\
  \subfloat[Posterior ensemble of $\tau_{yy}$]{\includegraphics[width=0.45\textwidth]{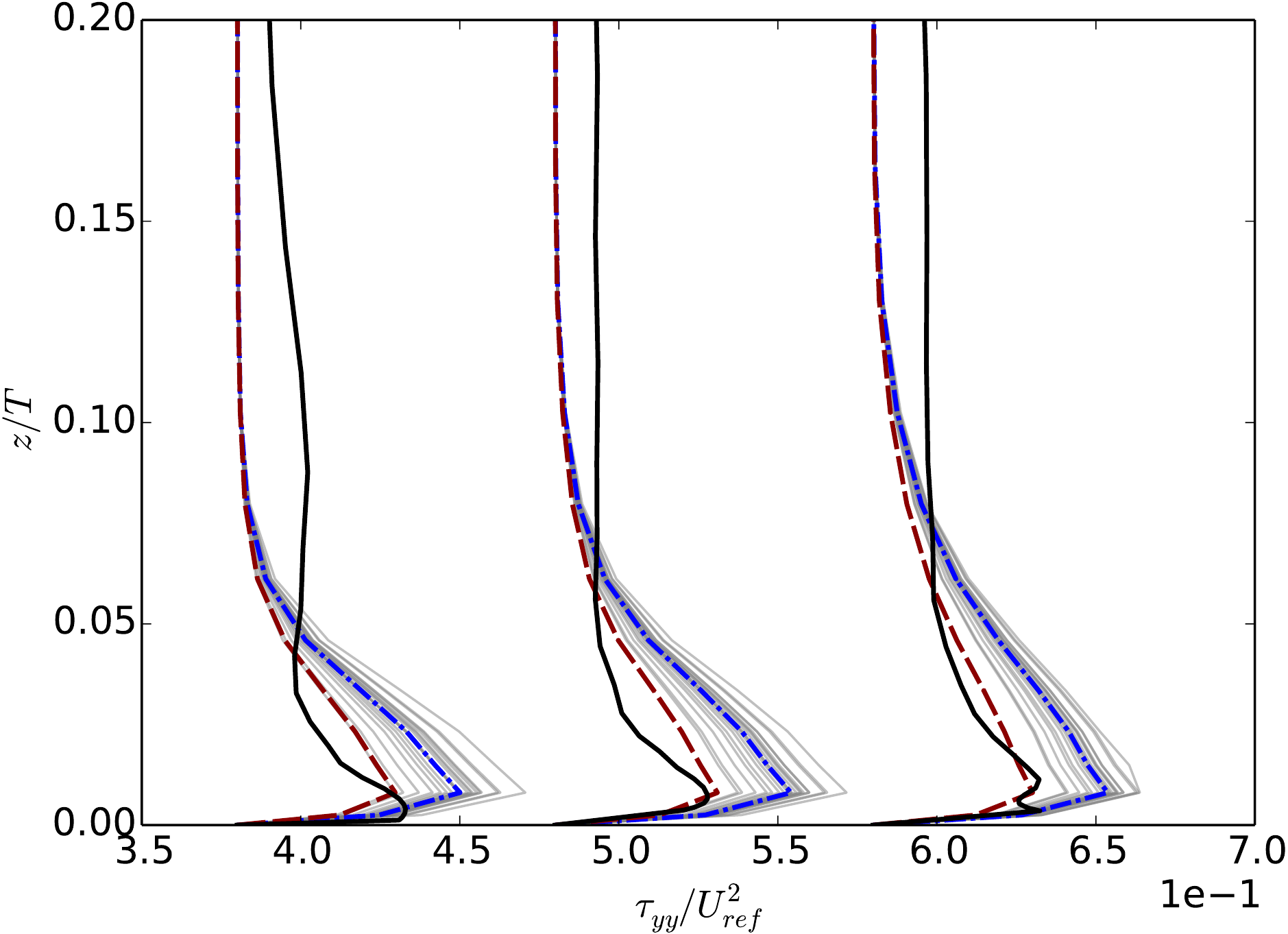}}\hspace{0.5em}
  \subfloat[Posterior ensemble of $\tau_{zz}$]{\includegraphics[width=0.45\textwidth]{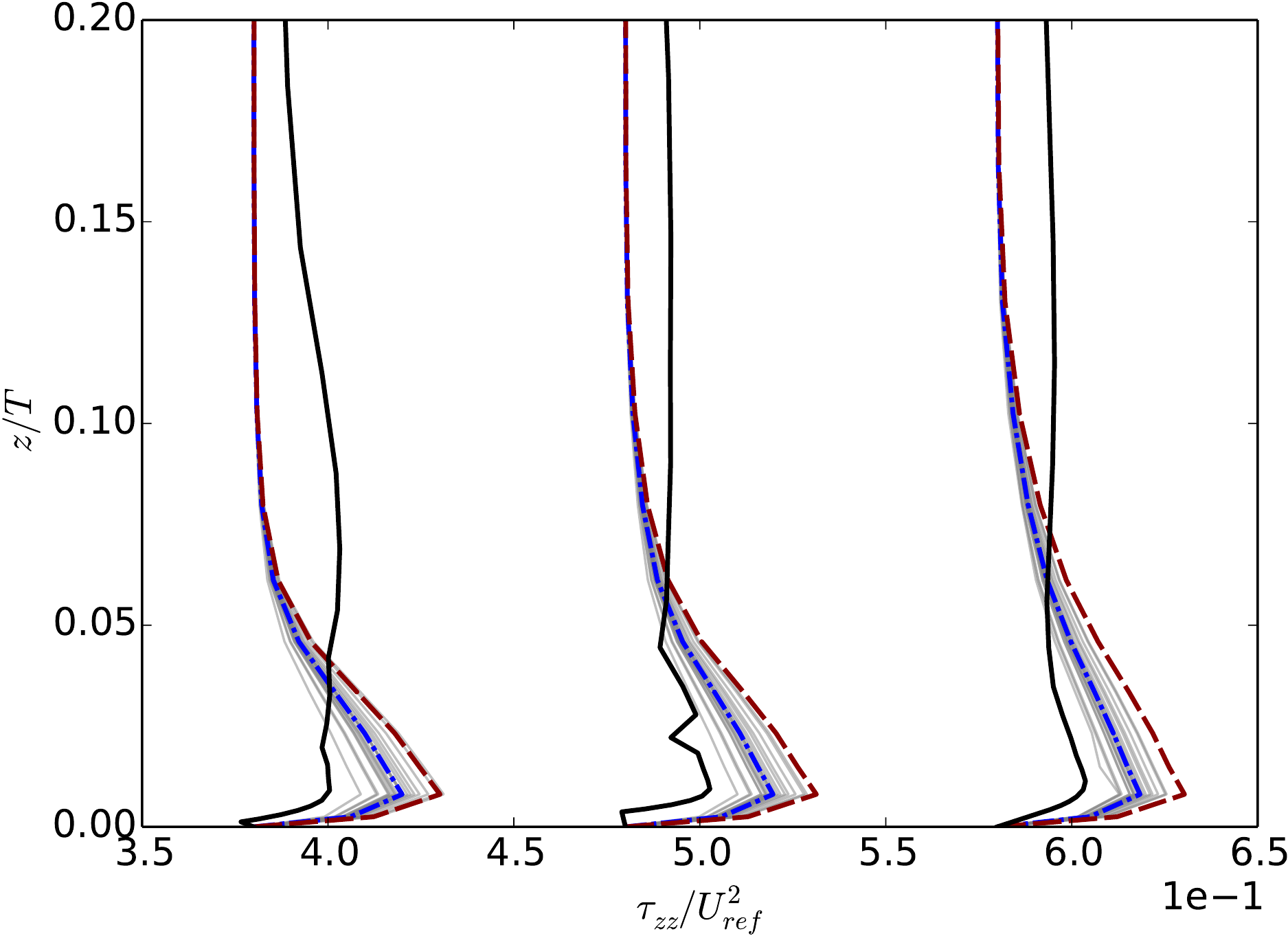}}\
  \caption{Posterior ensemble of Reynolds stress components at plane B ($x/T=3.187$) along three locations $-y/T=0.38$, $0.48$ and $0.58$, with increasing distance from the airfoil surface. A smaller value of $-y/T$ represents a location closer to the airfoil. Two components $\tau_{yy}$ and $\tau_{zz}$ are shown. The results of other components have the same trend and are omitted for simplicity. }
\label{fig:tau_comp_plane8}
\end{figure}

To examine the inference of Reynolds stress components, we use $\tau_{yy}$ and $\tau_{zz}$ as illustration. Figure~\ref{fig:tau_comp_plane8} shows these two components of Reynolds stress at plane B ($x/T=3.187$) along $-y/T=0.38$. It can be seen from Fig.~\ref{fig:tau_comp_plane8} that both components show little improvement. It is due to the fact that the mapping from velocity field to Reynolds stress field is not unique based on the RANS equations. Therefore, different sets of Reynolds stress field may satisfy the same velocity field, and the information incorporated by the velocity observation is not sufficient to infer each Reynolds stress component individually. 

\begin{figure}[htbp]
  \centering
   \hspace{1.5em}\includegraphics[width=0.6\textwidth]{U-legend-noObs}\\
  \includegraphics[width=0.6\textwidth]{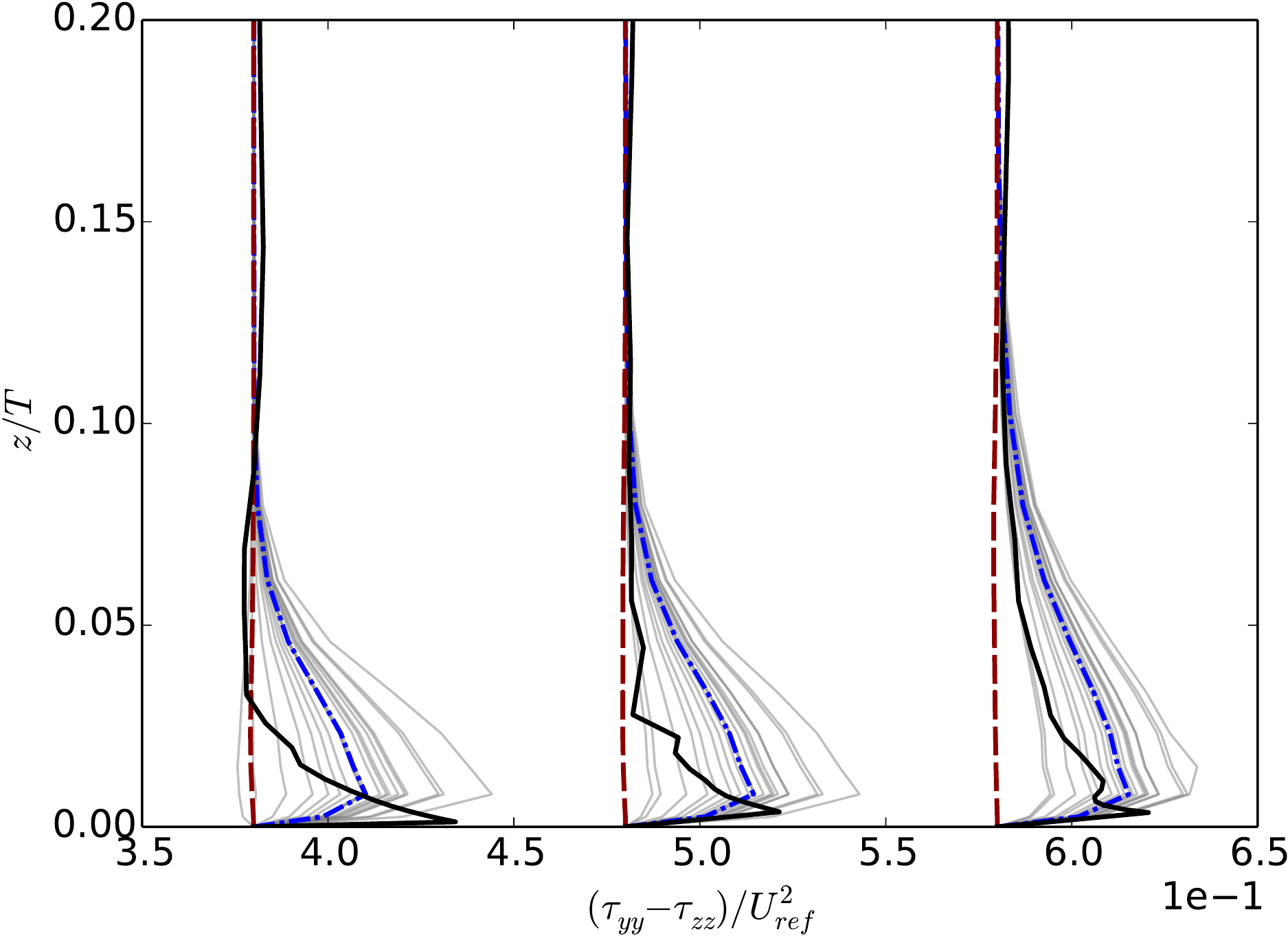}
  \caption{Posterior ensemble of normal stress imbalance $\tau_{yy}-\tau_{zz}$ in plane B ($x/T=3.187$) along three locations $-y/T=0.38$, $0.48$ and $0.58$, with increasing distance from the airfoil surface. A smaller value of $-y/T$ represents a location closer to the airfoil.. The normal stress imbalance is normalized with square of free stream velocity $U_{ref}^2$.}
  \label{fig:tau_vw_plane8}
\end{figure}

Unlike each component of Reynolds stress tensor, the normal stress imbalance $\tau_{yy}-\tau_{zz}$ is known to have direct impact upon the secondary flow at the corner region~\cite{huser1993direct}. The posterior ensemble of the normal stress imbalance $\tau_{yy}-\tau_{zz}$ is shown in Fig.~\ref{fig:tau_vw_plane8}. It can be seen that the baseline RANS prediction is close to zero due to the linear eddy viscosity assumption. According to the eddy viscosity assumption, the normal stress imbalance predicted by RANS simulation is related to the mean strain rates $\frac{\partial V}{\partial y}$ and $\frac{\partial W}{\partial z}$. These mean strain rates are close to zero based on the prediction of RANS models with linear eddy viscosity assumption. Consequently, the normal stress imbalance is close to zero and the stress-induced secondary flow is not captured by the baseline RANS simulation. Compared to the baseline RANS prediction, the posterior normal stress imbalance $\tau_{yy}-\tau_{zz}$ shows much better agreement with the experimental data, which is the main reason that the secondary flow prediction are improved by the Bayesian framework as shown in Fig.~\ref{fig:plane-8} and Fig.~\ref{fig:plane-10}.

\subsection{Results at Upstream of Leading Edge}
\begin{figure}[htbp]
  \centering
  \includegraphics[width=0.43\textwidth]{Tau-legend-Tri}\\
  \vspace{1em}
  \includegraphics[width=0.35\textwidth]{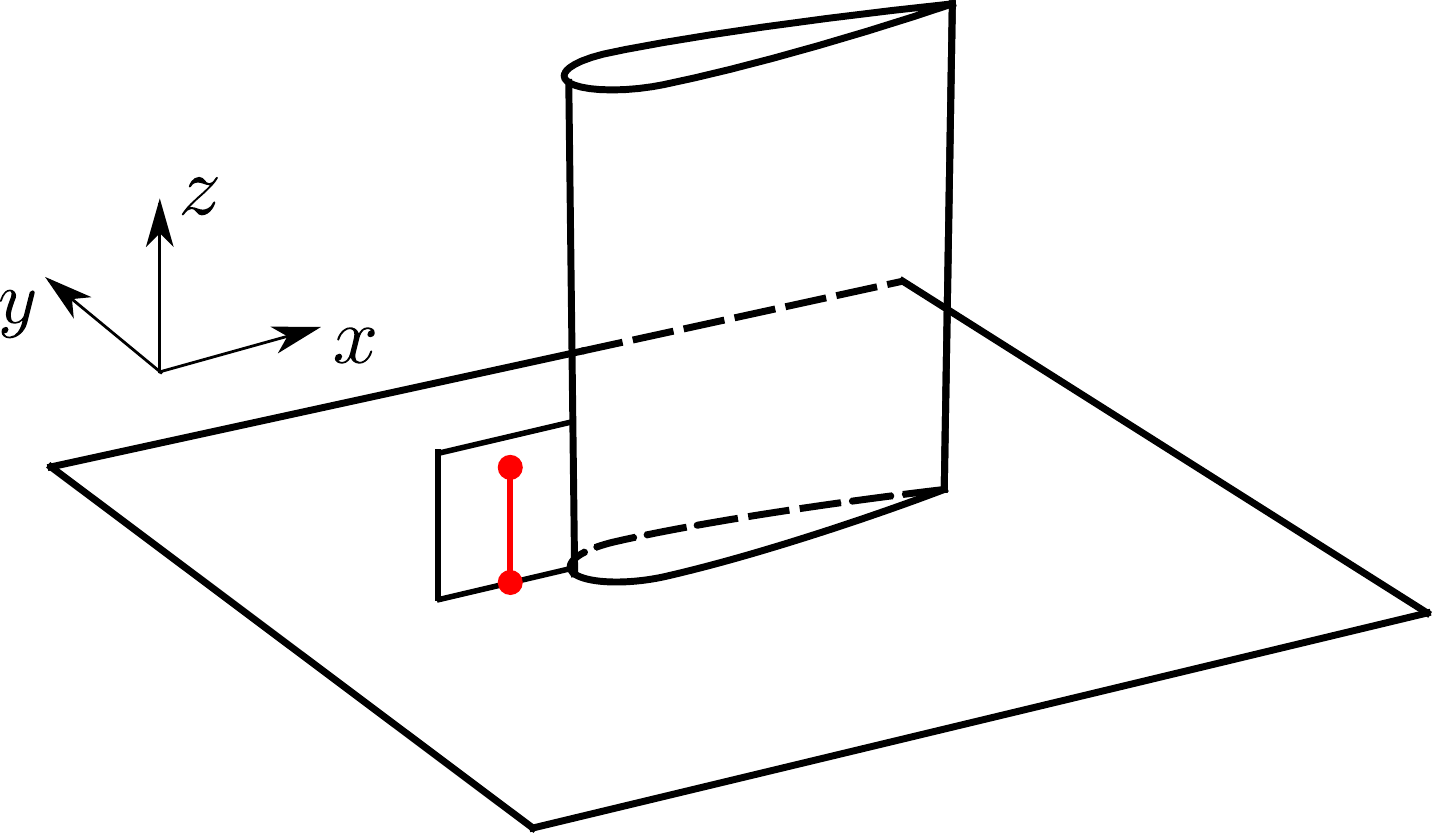}\\
  \subfloat[Prior]{\includegraphics[width=0.45\textwidth]{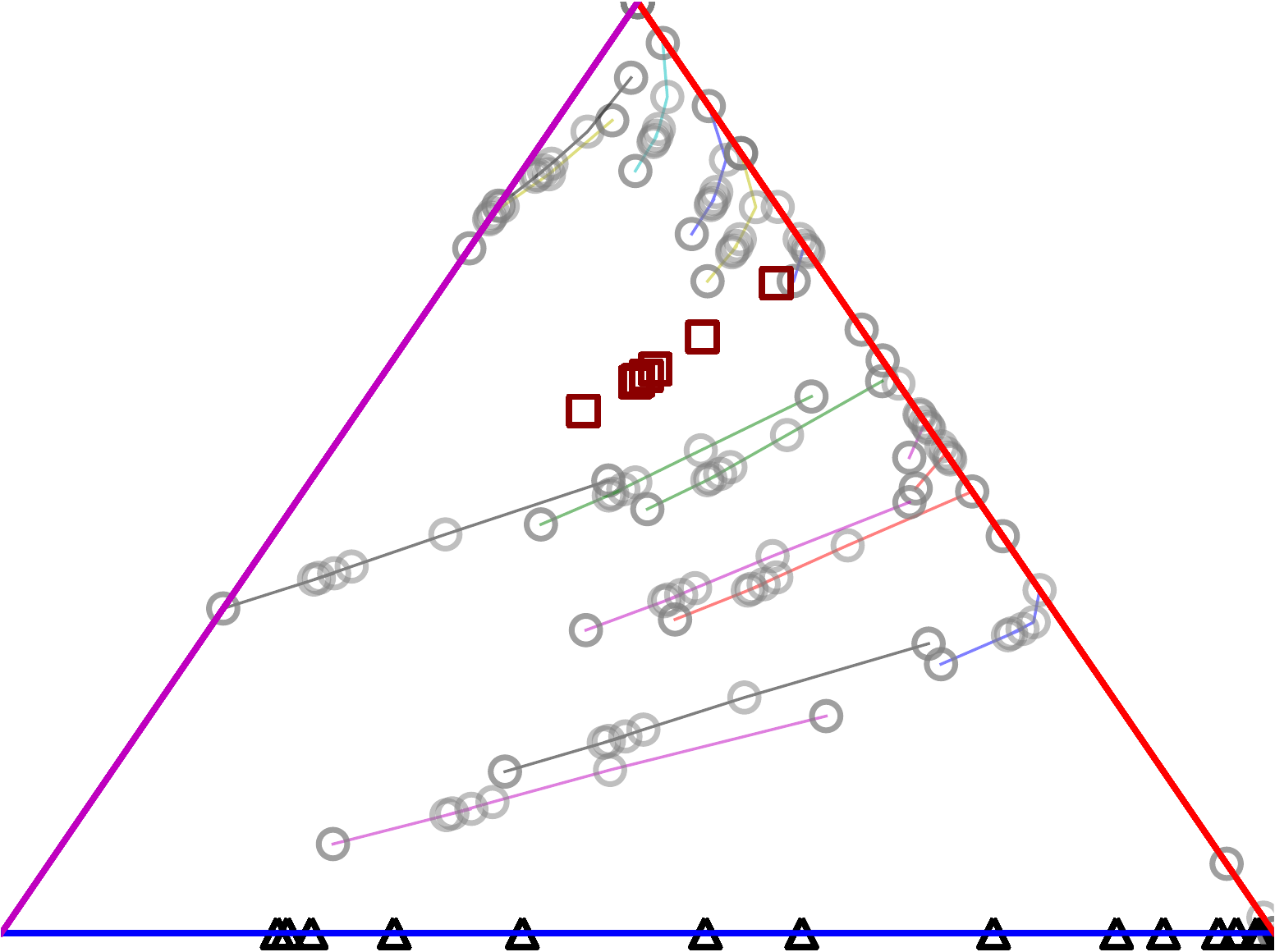}}\hspace{0.5em}
  \subfloat[Posterior]{\includegraphics[width=0.45\textwidth]{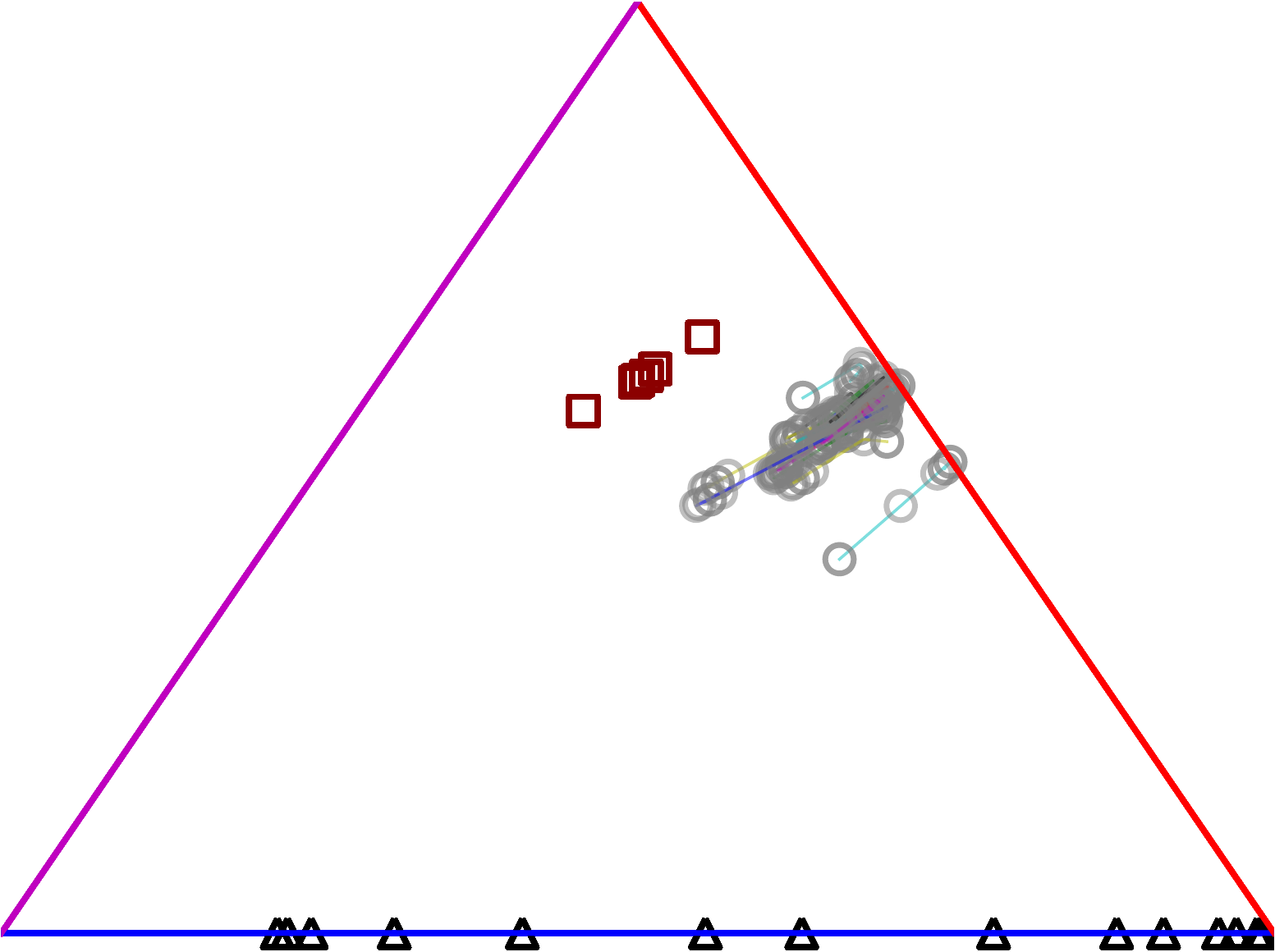}}
  \caption{Prior and posterior ensemble of Reynolds stresses in Barycentric triangle at $x/T$=-0.2. Only the Reynolds stresses at first 9 mesh points from the flat plate are shown for clarity. The experimental results fall on the bottom edge of the Barycentric triangle, which represents the two-component limit of turbulence.}
  \label{fig:tau-plane1}
\end{figure}

Figure~\ref{fig:tau-plane1}a shows the prior ensemble of Reynolds stresses upstream of the leading edge near $x/T=-0.2$, where the flow is approaching the stagnation point ($x/T$= 0) at the leading edge. Near this region, the horseshoe vortex develops when the incoming boundary layer experiences the strong adverse pressure gradient caused by the stagnation. According to the experimental measurement, the center of the horseshoe vortex system approximately locates at $x/T=-0.2$. Since the RANS models have difficulty in predicting the separation of the boundary layer, the RANS predicted Reynolds stresses are unreliable at this region and thus large perturbation is required. Based on this physics-informed knowledge, larger variance $\sigma$ is specified around this region. Consequently, the prior ensemble of Reynolds stresses shows a large scattering in the Barycentric triangle in Fig.~\ref{fig:tau-plane1}a. 

\begin{figure}[htbp]
  \centering
  \includegraphics[width=0.6\textwidth]{U-legend}\\
  \vspace{1em}
  \includegraphics[width=0.35\textwidth]{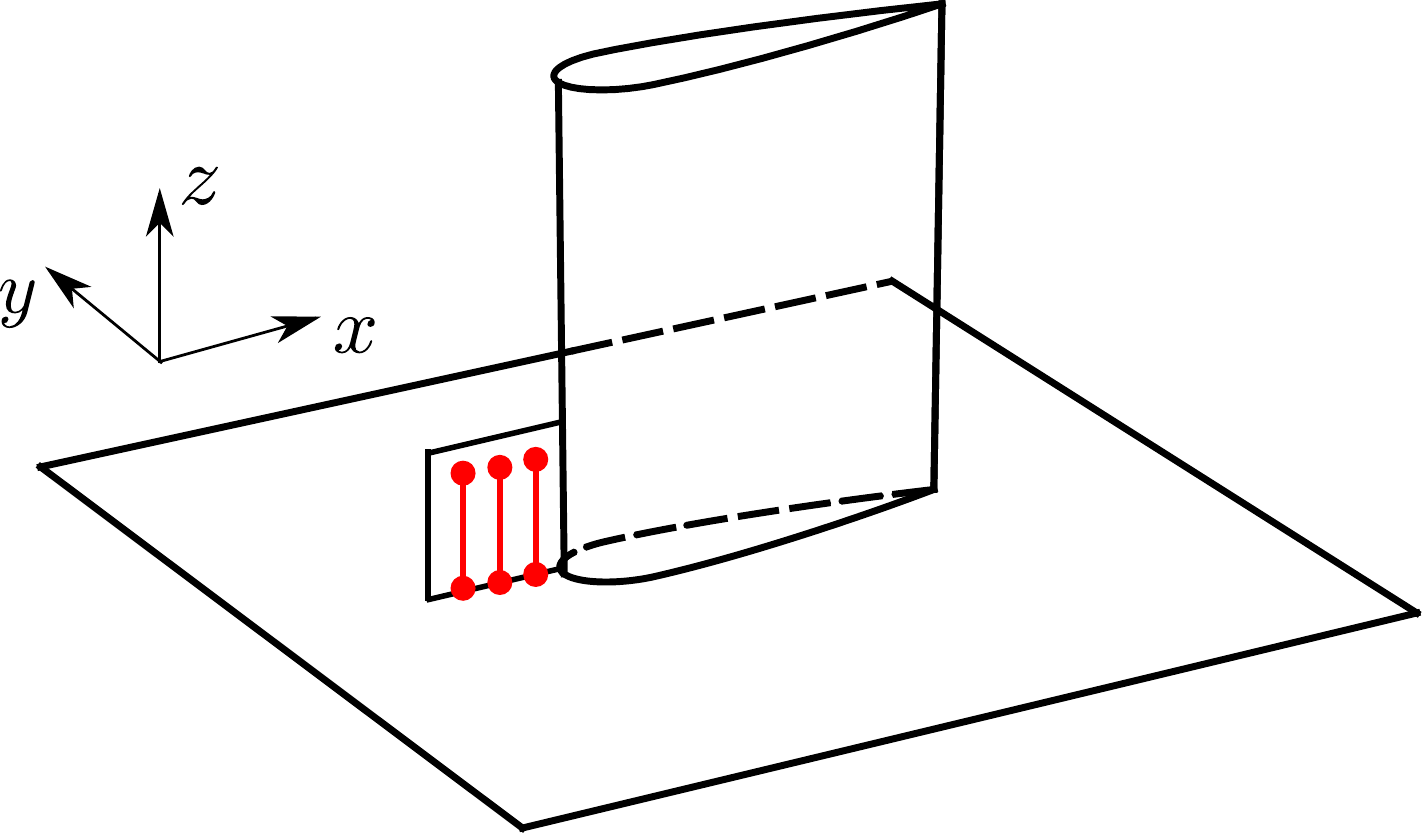}\\
  \subfloat[Prior]{\includegraphics[width=0.45\textwidth]{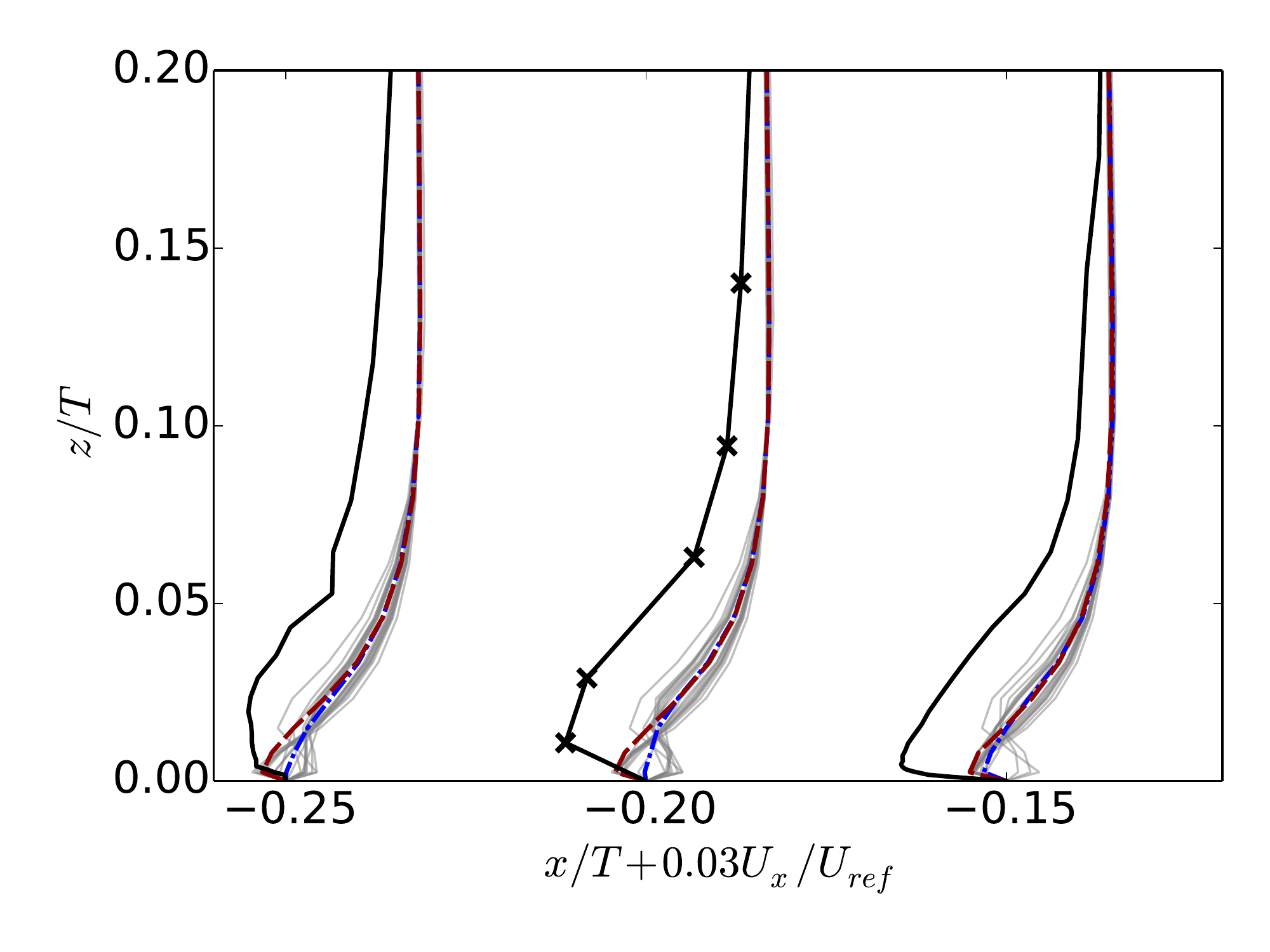}}\hspace{0.5em}
  \subfloat[Posterior]{\includegraphics[width=0.45\textwidth]{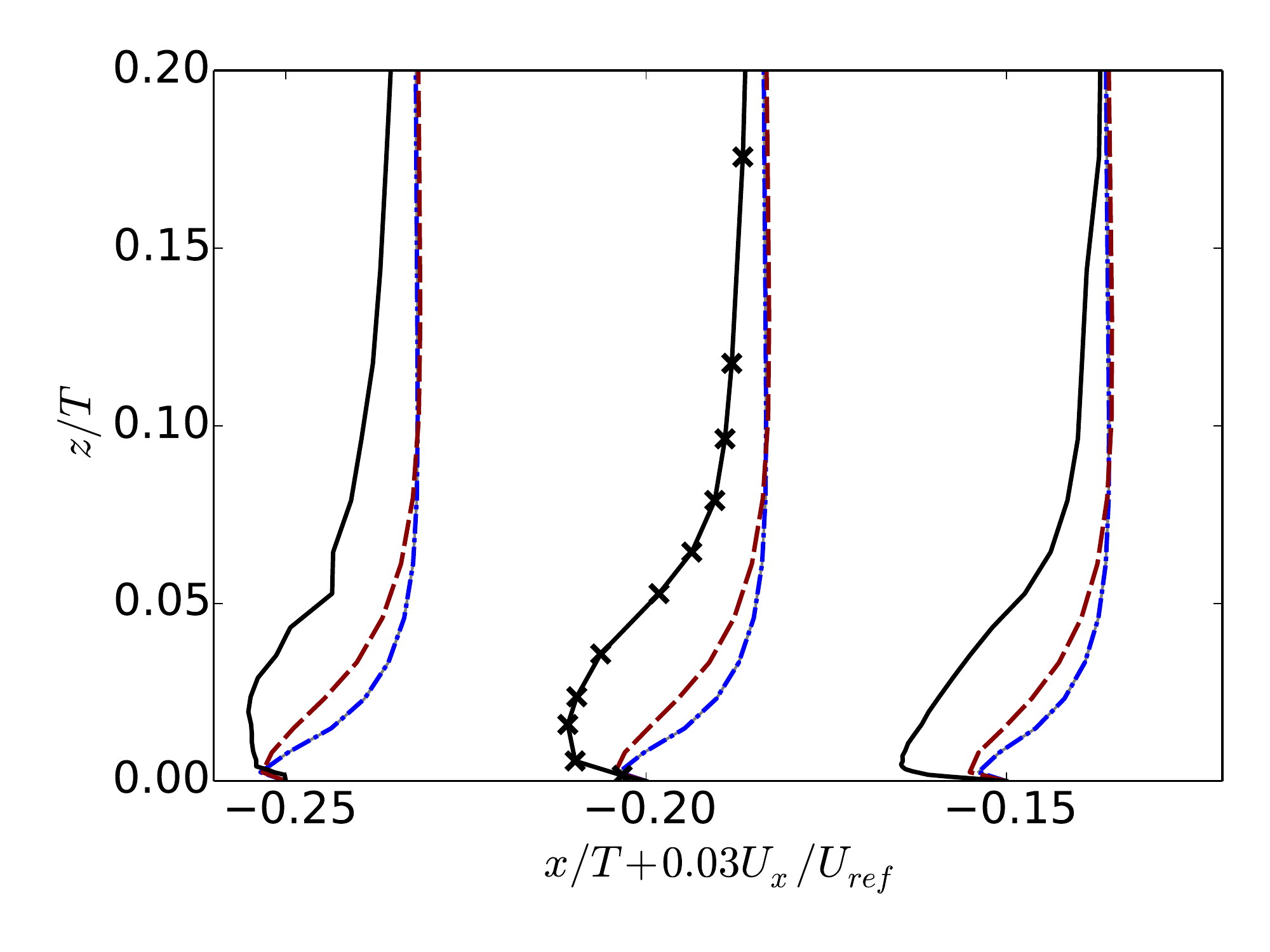}}
  \caption{The (a) prior and (b) posterior ensemble of streamwise velocity at the plane A upstream of the leading edge. The profiles are shown along three locations at $x/T=-0.25$, $-0.2$ and $-0.15$, with $x/T=-0.25$ being furtherest from the leading edge and $x/T=-0.15$ being closest.}
  \label{fig:Ux-plane1}
\end{figure}

The posterior ensemble of Reynolds stress is shown in Fig.~\ref{fig:tau-plane1}b. It can be seen that the inferred Reynolds stresses show no better agreement with the experimental data. Specifically, the experimental data shows that the Reynolds stress anisotropy starts from the one component limit state, since the Reynolds stress is restricted in two directions at the point that close to both the wall and the leading edge. For the sampled point further away from the plate, the Reynolds stress anisotropy becomes closer to the two component limit state, indicating that the restriction along the $z$ direction due to the wall is gradually reduced. We also examined the inferred velocity at this region. The prior and posterior velocity profiles at the plane upstream of the leading edge are shown in Figure~\ref{fig:Ux-plane1}. It can be seen that the posterior velocity profiles shown in Fig.~\ref{fig:Ux-plane1}b barely show better agreement with the experimental data compared to the RANS baseline prediction. Therefore, it indicates that the inference of Reynolds stress is not satisfactory and thus the propagated velocity field is not improved. In addition, it should be noted that the uncertainty as shown in the prior velocity profiles in Fig.~\ref{fig:Ux-plane1}a is not able to cover the experimental data, which indicates that the uncertainty space of Reynolds stress is restricted and may not cover the true Reynolds stress. Such restriction of uncertainty space of Reynolds stress can explain the unsatisfactory performance of the Bayesian inference at this region.

There are two possible reasons for the restriction of uncertainty space. First, the Reynolds stress anisotropy is not the dominant factor of the discrepancy of velocity field at this region. Since the turbulence time scale is known to be comparable with the time scale of the strain rate, the Reynolds stress is not able to adjust to the rapid change of strain rate~\cite{popebook}. Therefore, there is a misalignment of principal axis between the Reynolds stress tensor and strain rate tensor. However, the RANS models with linear eddy viscosity assumption predicts the Reynolds stress tensor that have the same principal axis as the strain rate tensor. Therefore, the principal axis of the Reynolds stress tensor predicted by RANS models has a large misalignment with the experimental data near the stagnation point, and the uncertainty space do not cover the true Reynolds stress since the orientation of the RANS predicted Reynolds stress is not perturbed in this framework. Consequently, the propagated velocity profiles barely show improvement compared to the RANS baseline prediction. Another reason is that the preservation of relative location of Reynolds stresses in Barycentric triangle as shown in Fig.~\ref{fig:tau-plane1}a. Such preservation of relative location has been discussed in Sec.~\ref{sec:result-corner}.

\begin{figure}[htbp]
  \centering
  \hspace{1.5em}\includegraphics[width=0.6\textwidth]{U-legend}\\
  \includegraphics[width=0.6\textwidth]{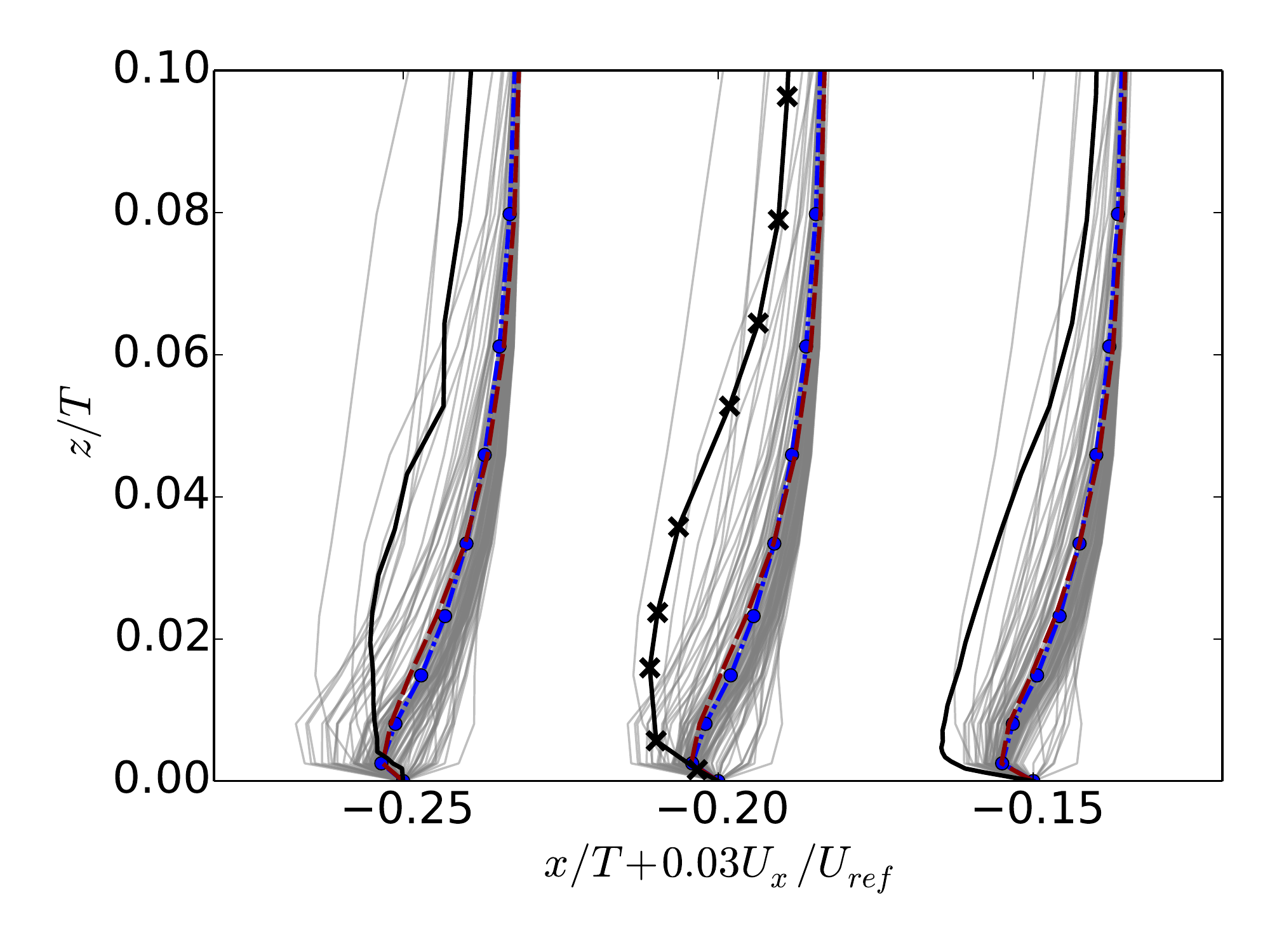}
  \caption{Streamwise velocity profiles at the plane upstream of the leading edge obtained from the prior Reynolds stresses, where both the Reynolds stress anisotropy and their orientations are perturbed. The profiles are shown along three locations at $x/T=-0.25$, $-0.2$ and $-0.15$ with $x/T=-0.25$ being furtherest  away from the leading edge and $x/T=-0.15$ being closest. }
  \label{fig:Ux-plane1-all}
\end{figure}

To further examine the dominant reason for the unsatisfactory inference performance at upstream of the leading edge, we first constructed another perturbation field via Gaussian process with smaller length scale at this region. The objective is to reduce the preservation of relative locations of Reynolds stresses and thus to reduce the restriction of the uncertainty space. By propagating the perturbed Reynolds stresses field to the velocity field, we find that both the prior and posterior velocity profiles are still similar to the ones as shown in Fig.~\ref{fig:Ux-plane1} and is omitted here for simplicity. It shows that the preservation of relative location of Reynolds stress in Barycentric triangle is not the dominant reason for the restriction of uncertainty space of velocity field near the stagnation point. Therefore, it is more likely that the misalignment of the principal axis of Reynolds stress is the main reason that accounts for the unsatisfactory inference performance at this region. Figure~\ref{fig:Ux-plane1-all} shows the prior velocity profiles near the stagnation point by enabling the perturbation of the orientation of Reynolds stresses. Compared to the velocity profiles as shown in Fig.~\ref{fig:Ux-plane1}a, the velocity profiles as shown in Fig.~\ref{fig:Ux-plane1-all} demonstrates that the velocity field at this region can cover the experimental data if the orientation of RANS predicted Reynolds stresses is perturbed. 

\begin{figure}[htbp]
  \centering
  \includegraphics[width=0.7\textwidth]{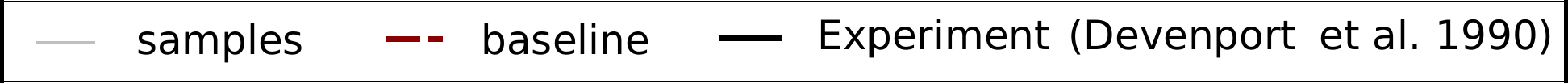}\\
  \vspace{1em}
  \includegraphics[width=0.35\textwidth]{rood-3D-Plane1-0p2}\\
  \subfloat[$\mathbf{v}_1$ at $z/T=0.035$]{\includegraphics[width=0.27\textwidth]{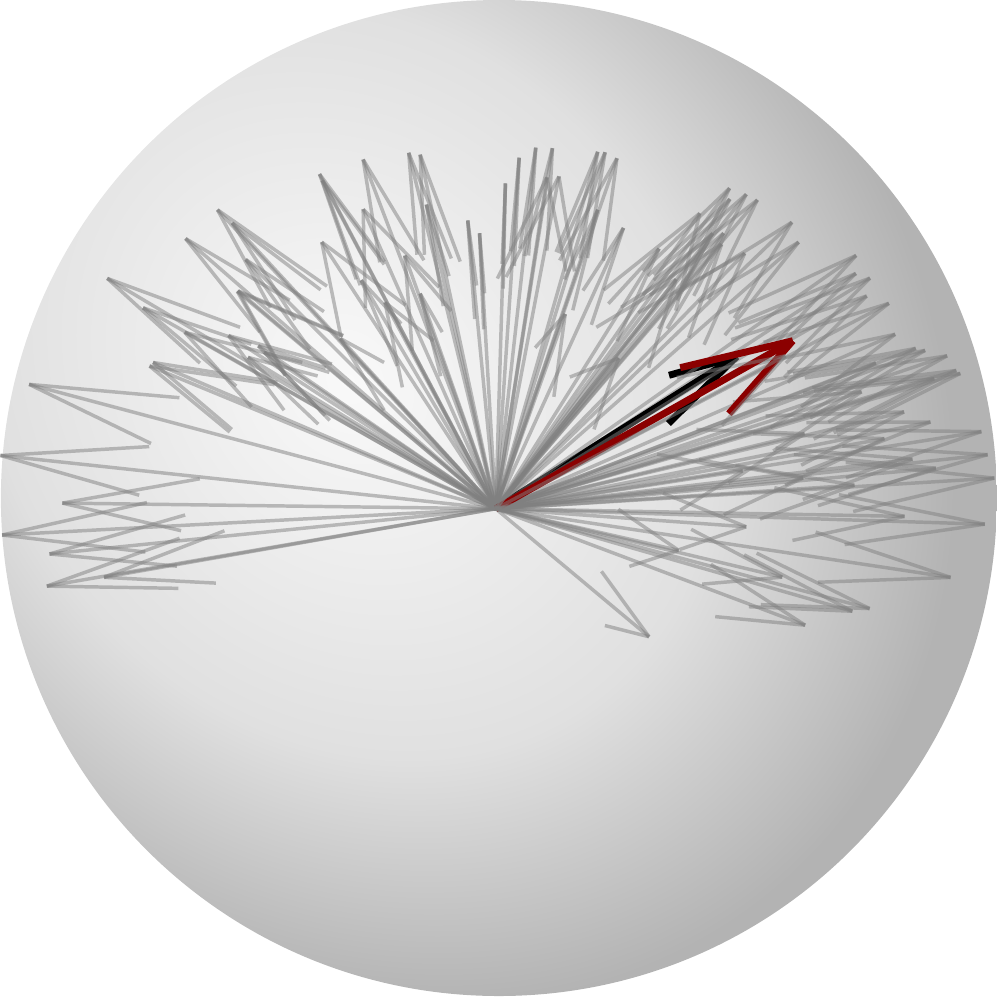}}\hspace{2em}
  \subfloat[$\mathbf{v}_1$ at $z/T=0.015$]{\includegraphics[width=0.27\textwidth]{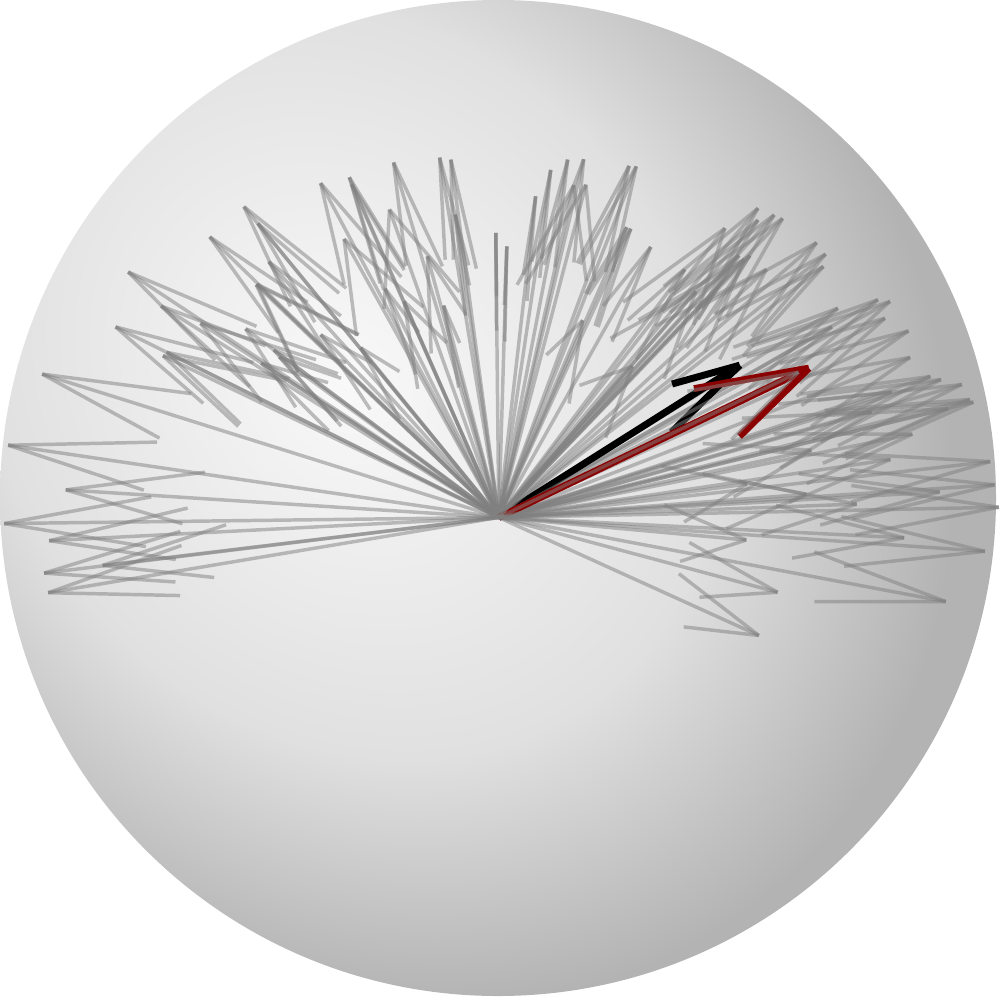}}\hspace{2em}
  \subfloat[$\mathbf{v}_1$ at $z/T=2.5 \times 10^{-3}$]{\includegraphics[width=0.27\textwidth]{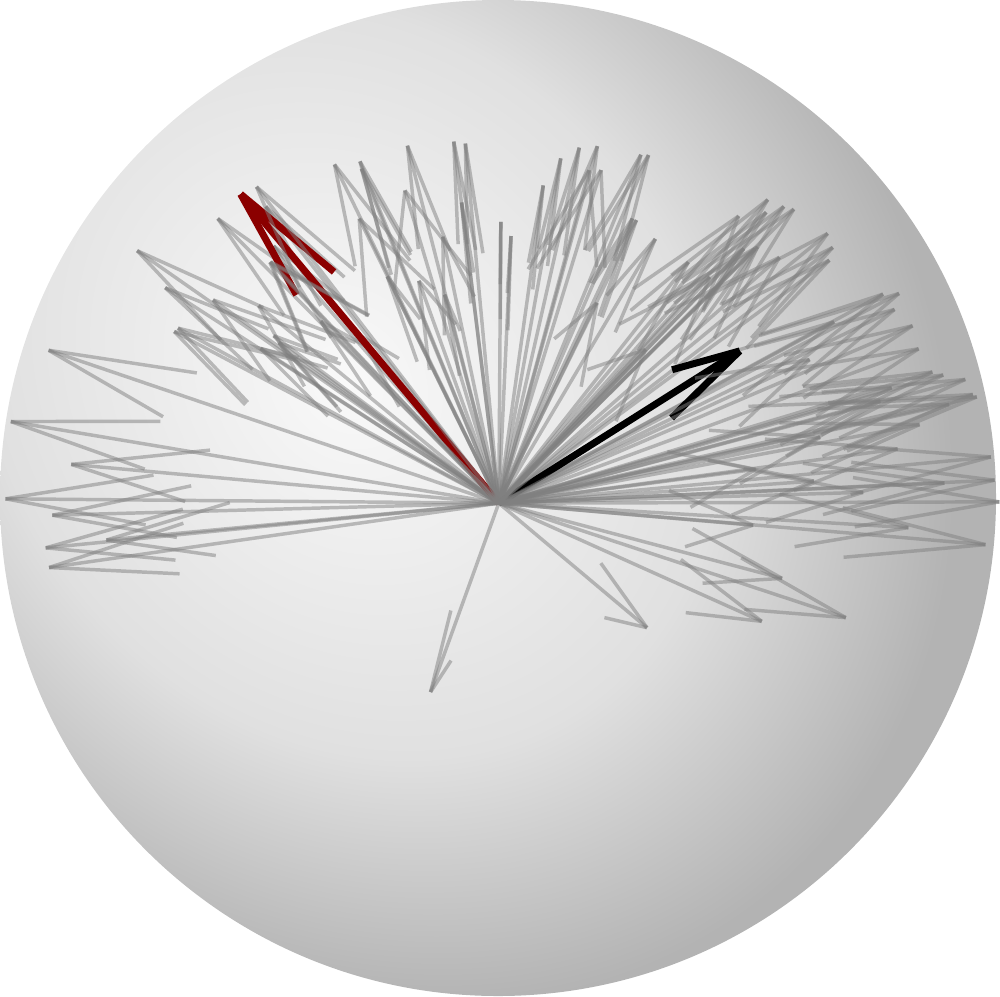}}\hspace{2em}
  \subfloat[$\mathbf{v}_2$ at $z/T=0.035$]{\includegraphics[width=0.26\textwidth]{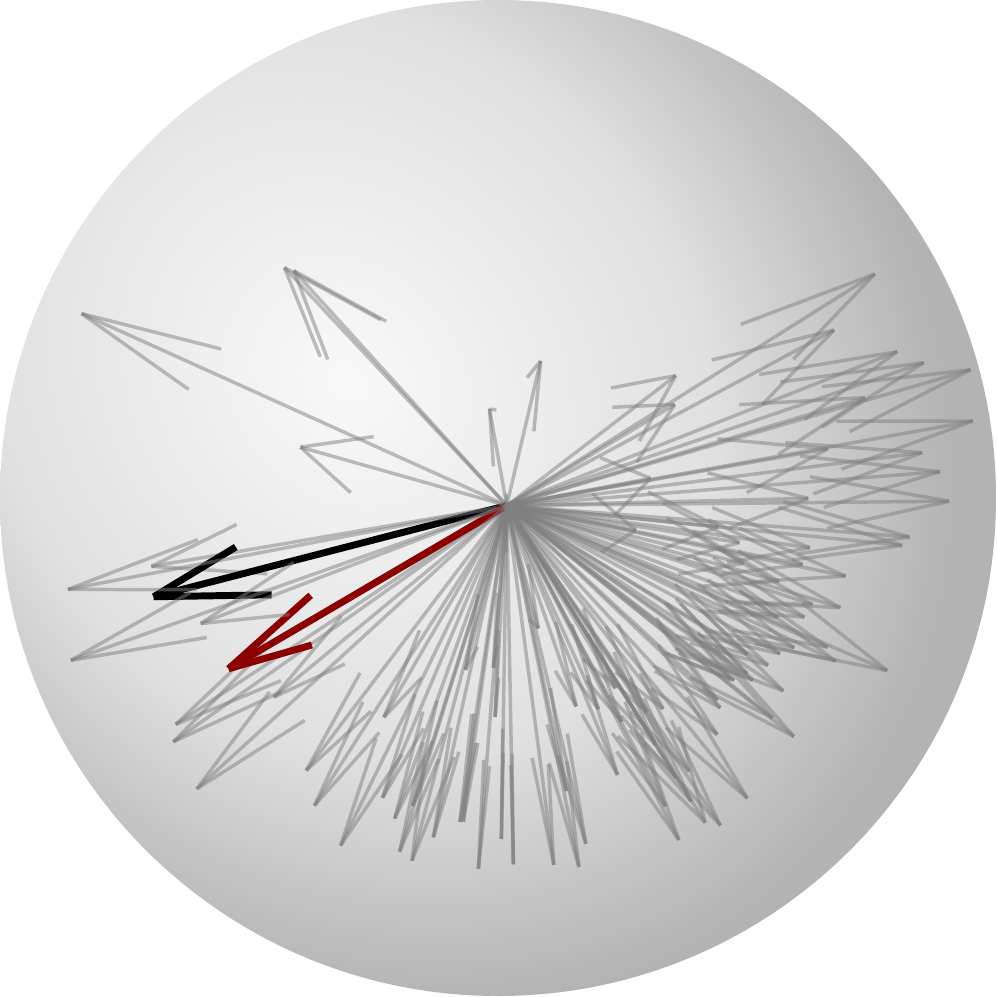}} \hspace{2em}
  \subfloat[$\mathbf{v}_2$ at $z/T=0.015$]{\includegraphics[width=0.26\textwidth]{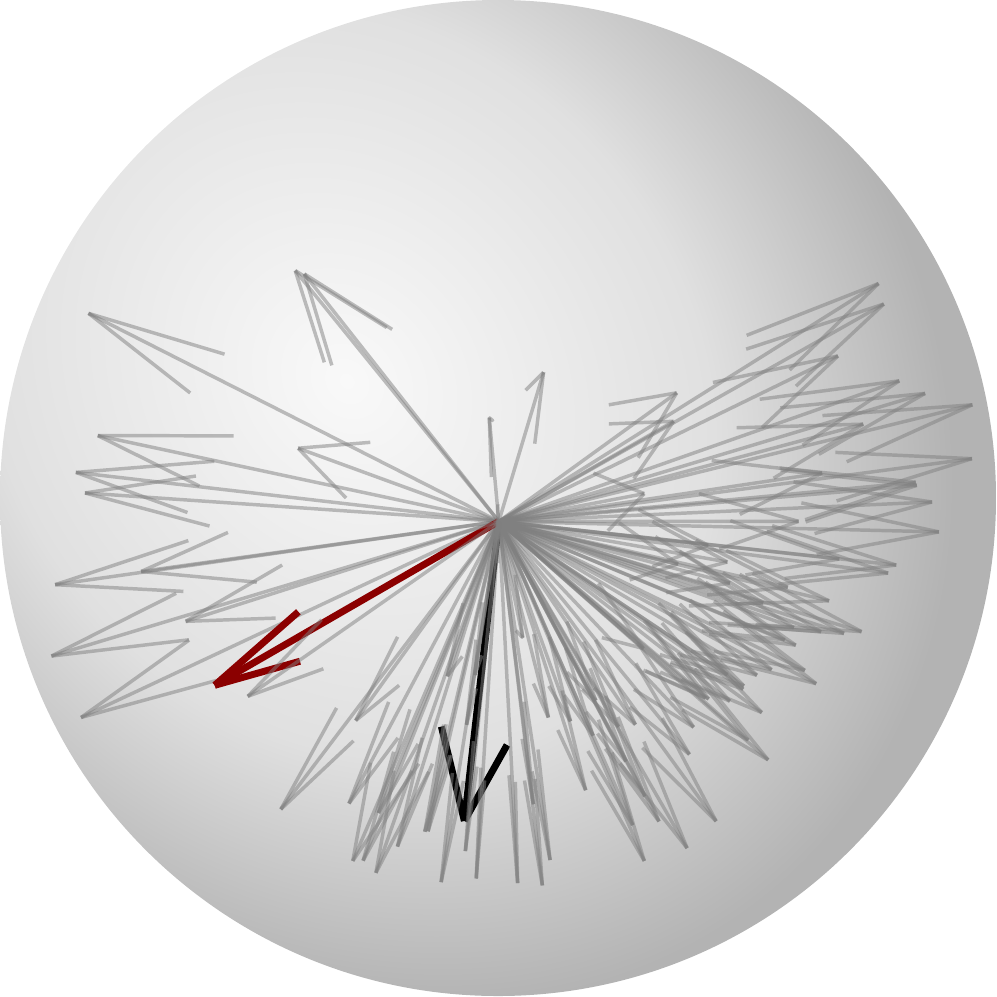}} \hspace{2em}
  \subfloat[$\mathbf{v}_2$ at $z/T=2.5 \times 10^{-3}$]{\includegraphics[width=0.26\textwidth]{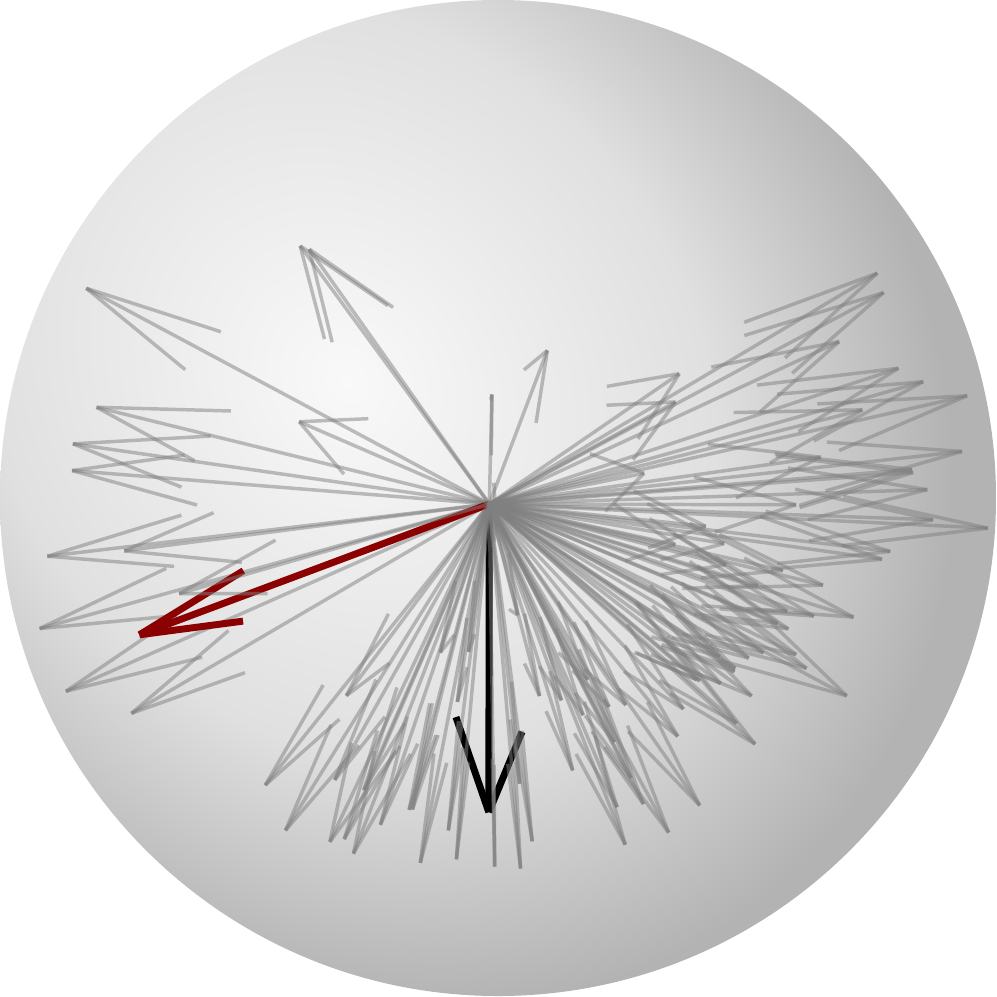}}
  \caption{The comparison of orientations (as indicated by the normal eigenvectors $\mathbf{v}_1$
    and $\mathbf{v}_2$) of the Reynolds stresses. The Reynolds stresses are sampled at three
    locations with vertical coordinates $z/T=0.035$ (panels a and d), $0.015$ (panels b and e) and
    $2.5 \times 10^{-3}$ (panels c and f).  The third point $z/T=2.5 \times 10^{-3}$ is located at
    the center of the first cell next to the bottom flat plate. All three points are located along
    the line $x/T=-0.2$ upstream the leading edge (indicated below the legend).  The orientation
    vector $\mathbf{v}_3$ is omitted for clarity since it can be uniquely determined from
    $\mathbf{v}_1$ and $\mathbf{v}_2$, i.e., $\mathbf{v}_3 = \mathbf{v}_1 \times \mathbf{v}_2$.}
  \label{fig:orient}
\end{figure}

To illustrate the misalignment of principal axis, the RANS predicted orientation of Reynolds stress
($\mathbf{v}_1$, $\mathbf{v}_2$ and $\mathbf{v}_3$) and the experimental data are shown in
Fig.~\ref{fig:orient}. The perturbed orientations of Reynolds stress, which correspond to the prior
velocity profiles as shown in Fig.~\ref{fig:Ux-plane1-all}, are also shown in
Fig.~\ref{fig:orient}. The Reynolds stresses are sampled at three locations, one of which is located
at the first cell near the flat plate ($z/T=2.5\times 10^{-3}$), and the other two are further away
from the plate at $z/T=0.015$ and $z/T=0.035$, respectively.The unit vectors as shown in
Fig.~\ref{fig:orient} represents the direction of principal axis. It can be seen from
Fig.~\ref{fig:orient} that the misalignment of principal axis between RANS predicted Reynolds stress
and the experimental data is more severe at the region near the flat plate. With the increase of the
distance away from the plate, the misalignment of principal axis is reduced, especially for the
orientation of $\mathbf{v}_1$. By injecting uncertainties into the orientation of Reynolds stress,
it can be seen in Fig.~\ref{fig:orient} that the samples of Reynolds stress orientation show a
scattering around the baseline RANS result and cover the experimental data. It explains the better
coverage of velocity profiles as shown in Fig.~\ref{fig:Ux-plane1-all}. In addition, it demonstrates
that enabling the perturbation of the orientation $\mathbf{v}_1$, $\mathbf{v}_2$, $\mathbf{v}_3$ can
better explore the uncertainty space of Reynolds stress and indicates that the inference performance
at the leading edge can be improved. However, it should be noted that the perturbation of the
orientation $\mathbf{v}_1$, $\mathbf{v}_2$, $\mathbf{v}_3$ can lead to the momentum flux from low
momentum cell to high momentum one, which can cause numerical instability.  Moreover, given the same
amount of observation data, introducing more unknowns into the problem increases the dimensionality
of the uncertainty space and thus inevitably increases the difficulty for the inference.  Extensions
of our current Bayesian inference framework to include perturbed orientations and applications to the wing-body
junction flow problem are under way.

\section{Conclusion}
\label{sec:conclude}

In this work we apply the Bayesian framework proposed by Xiao et al.~\cite{xiao-mfu} to the wing--body junction flow, which is featured by the horseshoe vortex system and the possible corner separation. Both features are studied in this work to evaluate the performance of the Bayesian framework for the complex flow problems. To reduce the computational cost, we extend the original Bayesian framework to account for the RANS simulations with wall function. Simulation results suggest that, at the corner region, both the posterior mean velocities and the Reynolds stress anisotropy show better agreement with the experimental data, even though the posterior Reynolds stress components do not demonstrate noticeable improvement. The improvement in the prediction of mean flow field and Reynolds stress anisotropy demonstrate the capability of this Bayesian framework in predicting the complex flow problem. In addition, the analysis of posterior Reynolds stress components indicates that the secondary flow at corner region is largely governed by the Reynolds stress anisotropy, and it is not necessary to accurately model each component of Reynolds stress to improve the mean flow field prediction at this region.

On the other hand, at the center region of horseshoe vortex, the prior mean velocities are still
close to the baseline RANS prediction, even though the prior Reynolds stress anisotropy shows a
significant difference from the baseline predicted Reynolds stress in Barycentric triangle. In
addition, the posterior mean velocities at leading edge show little improvement compared with the
experimental data. These results indicate that the mean flow field at this region is not as
sensitive to the Reynolds stress anisotropy as the flow at the corner region. This is attributed to
the rapid change of strain rate near the stagnation point, which leads to a misalignment of
principal axis between RANS predicted Reynolds stresses and the experimental data. In this work, the
orientations of RANS predicted Reynolds stress are not perturbed, and thus the uncertainty space would
not cover the experimental data. It explains the unsatisfactory inference performance near the
leading edge where the strain rate changes rapidly as the flow is approaching the stagnation
point. In order to account for the flow problem with such rapid change of mean strain rate, consideration of
the uncertainties in the orientations of Reynolds stress is necessary to extend the capability of
the current framework.

\bibliographystyle{elsarticle-num}

\begin{thebibliography}{10}
\expandafter\ifx\csname url\endcsname\relax
  \def\url#1{\texttt{#1}}\fi
\expandafter\ifx\csname urlprefix\endcsname\relax\def\urlprefix{URL }\fi
\expandafter\ifx\csname href\endcsname\relax
  \def\href#1#2{#2} \def\path#1{#1}\fi

\bibitem{devenport1990}
W.~J. Devenport, R.~L. Simpson, Time-depeiident and time-averaged turbulence
  structure near the nose of a wing-body junction, Journal of Fluid Mechanics
  210 (1990) 23--55.

\bibitem{simpson2001}
R.~L. Simpson, Junction flows, Annual Review of Fluid Mechanics 33~(1) (2001)
  415--443.

\bibitem{rodi1998}
W.~Rodi, J.~Bonnin, T.~Buchal, D.~Laurence, Testing of calculation methods for
  turbulent flows: Workshop results for 5 test cases, Electricit{\'e} de France
  Report.

\bibitem{coombs2012}
J.~L. Coombs, C.~J. Doolan, D.~J. Moreau, A.~Zander, L.~Brooks, Assessment of
  turbulence models for a wing-in-junction flow, Assessment 3 (2012) 7.

\bibitem{apsley2001}
D.~Apsley, M.~Leschziner, Investigation of advanced turbulence models for the
  flow in a generic wing-body junction, Flow, Turbulence and Combustion 67~(1)
  (2001) 25--55.

\bibitem{chen1995}
H.-C. Chen, Assessment of a {Reynolds} stress closure model for appendage-hull
  junction flows, Journal of Fluids Engineering 117~(4) (1995) 557--563.

\bibitem{jones2005}
D.~A. Jones, D.~B. Clarke, Simulation of a wing-body junction experiment using
  the fluent code, Tech. rep., DTIC Document (2005).

\bibitem{alin2008}
N.~Alin, C.~Fureby, Large eddy simulation of junction vortex flows, in: 46th
  AIAA Aerospace Sciences Meeting \& Exhibit, 2008, pp. 7--10.

\bibitem{paik2007}
J.~Paik, C.~Escauriaza, F.~Sotiropoulos, On the bimodal dynamics of the
  turbulent horseshoe vortex system in a wing-body junction, Physics of Fluids
  (1994-present) 19~(4) (2007) 045107.

\bibitem{huser1993direct}
A.~Huser, S.~Biringen, Direct numerical simulation of turbulent flow in a
  square duct, Journal of Fluid Mechanics 257 (1993) 65--95.

\bibitem{gand2012}
F.~Gand, V.~Brunet, S.~Deck, Experimental and numerical investigation of a
  wing-body junction flow, AIAA journal 50~(12) (2012) 2711--2719.

\bibitem{gand2015}
F.~Gand, J.-C. Monnier, J.-M. Deluc, A.~Choffat, Experimental study of the
  corner flow separation on a simplified junction, AIAA Journal 53~(10) (2015)
  2869--2877.

\bibitem{bordji2014}
M.~Bordji, F.~Gand, V.~Brunet, S.~Deck, Comparative study of linear and
  non-linear {RANS} closures for corner flows, in: 32nd AIAA Applied
  Aerodynamics Conference, AIAA Aviation and Aeronautics Forum and Exposition
  2014, 2014.

\bibitem{rumsey2016}
C.~L. Rumsey, D.~Neuhart, M.~A. Kegerise, The {NASA} juncture flow experiment:
  Goals, progress, and preliminary testing, in: 54th AIAA Aerospace Sciences
  Meeting, 2016, p. 1557.

\bibitem{xiao-mfu}
H.~Xiao, J.-L. Wu, J.-X. Wang, R.~Sun, C.~J. Roy, Quantifying and reducing
  model-form uncertainties in {Reynolds-Averaged} {Navier-Stokes} simulations:
  An open-box, physics-based, bayesian approach, submitted. Available at
  http://arxiv.org/abs/1508.06315 (2015).

\bibitem{iglesias2013ensemble}
M.~A. Iglesias, K.~J. Law, A.~M. Stuart, Ensemble {Kalman} methods for inverse
  problems, Inverse Problems 29~(4) (2013) 045001.

\bibitem{oliver2011bayesian}
T.~A. Oliver, R.~D. Moser, Bayesian uncertainty quantification applied to rans
  turbulence models, in: Journal of Physics: Conference Series, Vol. 318, IOP
  Publishing, 2011, p. 042032.

\bibitem{tennekes1972first}
H.~Tennekes, J.~L. Lumley, A first course in turbulence, MIT press, 1972.

\bibitem{gorle2013framework}
C.~Gorl{\'e}, G.~Iaccarino, A framework for epistemic uncertainty
  quantification of turbulent scalar flux models for {Reynolds}-averaged
  {Navier}-{Stokes} simulations, Physics of Fluids 25~(5) (2013) 055105.

\bibitem{gorle2014deviation}
C.~Gorl{\'e}, J.~Larsson, M.~Emory, G.~Iaccarino, The deviation from parallel
  shear flow as an indicator of linear eddy-viscosity model inaccuracy, Physics
  of Fluids 26~(5) (2014) 051702.

\bibitem{emory2013modeling}
M.~Emory, J.~Larsson, G.~Iaccarino, Modeling of structural uncertainties in
  {Reynolds}-averaged {Navier}-{Stokes} closures, Physics of Fluids 25~(11)
  (2013) 110822.

\bibitem{emory2011modeling}
M.~Emory, R.~Pecnik, G.~Iaccarino, Modeling structural uncertainties in
  {Reynolds-averaged} computations of shock/boundary layer interactions, AIAA
  paper 479 (2011) 1--16.

\bibitem{emory14estimate}
M.~A. Emory, Estimating model-form uncertainty in {Reynolds-averaged}
  {Navier--Stokes} closures, Ph.D. thesis, Stanford University (2014).

\bibitem{banerjee2007presentation}
S.~Banerjee, R.~Krahl, F.~Durst, C.~Zenger, Presentation of anisotropy
  properties of turbulence, invariants versus eigenvalue approaches, Journal of
  Turbulence 8~(32) (2007) 1--27.

\bibitem{wu2015bayesian}
J.-L. Wu, J.-X. Wang, H.~Xiao, A {Bayesian} calibration-prediction method for
  reducing model-form uncertainties with application in {RANS} simulations,
  submitted. Available at http://arxiv.org/abs/1510.06040 (2015).

\bibitem{le2010spectral}
O.~P. Le~Ma{\^\i}tre, O.~M. Knio, Spectral methods for uncertainty
  quantification: with applications to computational fluid dynamics, Springer,
  2010.

\bibitem{evensen2009data}
G.~Evensen, Data assimilation: the ensemble {Kalman} filter, Springer, 2009.

\bibitem{popebook}
S.~B. Pope, Turbulent Flows, Cambridge University Press, Cambridge, 2000.

\bibitem{launder1974numerical}
B.~E. Launder, D.~Spalding, The numerical computation of turbulent flows,
  Computer methods in applied mechanics and engineering 3~(2) (1974) 269--289.

\end{thebibliography}

\end{document}